\documentclass[11pt,a4paper]{article}
 
\usepackage{float}
\usepackage{multirow}
\usepackage{graphics}
\usepackage[cp1252,utf8]{inputenc}
\usepackage{hyperref}
\usepackage{graphicx} 
\usepackage{amsmath}
\usepackage{amsfonts}
\usepackage{amssymb}
\usepackage{bigints} 
\usepackage{relsize} 
\usepackage[top=2cm, bottom=2.5cm, left=1.5cm, right=1.5cm]{geometry}
\usepackage{multicol}
\usepackage{authblk}
\usepackage{changepage} 
\usepackage{afterpage}
\usepackage{inputenc}
\usepackage{placeins}

\usepackage{graphicx}
\usepackage{bm}
\usepackage{hyperref}
\usepackage{mathrsfs}
\usepackage{subfig}
\usepackage{caption}

\usepackage[english]{babel}

\title{\bf Study of circular geodesics and shadow of rotating charged black hole surrounded by perfect fluid dark matter immersed in plasma}
\author[a]{\bf Anish Das\thanks{anishdasslg@gmail.com,anishdas1995@bose.res.in}}
\author[b]{\bf  Ashis Saha \thanks{sahaashis0007@gmail.com, ashisphys18@klyuniv.ac.in}}
\author[c]{\bf Sunandan Gangopadhyay \thanks{sunandan.gangopadhyay@gmail.com, sunandan.gangopadhyay@bose.res.in}}

\affil[a,c]{\textit{Department of Theoretical Sciences, S.N.~Bose National Centre for Basic Sciences,}
	\textit{JD Block, Sector-III, Salt Lake, Kolkata 700106, India}}

\affil[b]{\textit{Department of Physics, University of Kalyani, Kalyani 741235, India}}

\date{}

\begin{document}
\maketitle

\begin{abstract}

\noindent In this work, we consider a rotating charged black hole surrounded by perfect fluid dark matter. We consider the system to be immersed in non-magnetised, pressureless plasma. First, we evaluate the null geodesics in order to study the co-rotating and counter rotating photon orbits. Further, we analyse the null geodesics to calculate the celestial coordinates ($\alpha, \beta$). The celestial coordinates are used to determine the black hole shadow radius ($R_s$). Thereafter, we observe and analyse the effects of black hole spacetime, perfect fluid dark matter and plasma parameters ($a$, $Q$, $\chi$, $k$) on the black hole shadow in detail. Finally, we study the effect of plasma distribution on the effective potential ($V_{eff}$) of the black hole spacetime as encountered by the photons. We also present bounds on the plasma parameter from the observational data from $M87^{*}$ central supermassive black hole.

\end{abstract}

\maketitle

\section{Introduction}\label{sec1}
Recent advancement in observational astrophysics paved the way for the release of the first image of supermassive black hole \cite{2}. Black holes are one of the most intriguing predictions of general theory of relativity. It is predicted that supermassive black holes form the core of almost all the galaxies in the universe. The recent image released by the Event Horizon Telescope (\textit{EHT}) group depict the shadow of the supermassive black hole residing at the center of Messier 87 galaxy \cite{2}. Theoretical analysis of the formation of the image of black holes (\textit{black hole shadow}) started with the works of Synge \cite{3} who demonstrated the shadow of a spherically symmetric black hole. Later, Luminet \cite{4} studied the shadow of a spherically symmetric black hole surrounded by accretion disks. Shadows are dark disks formed by light coming from a background source when a black hole lies between the source and observer with no source between the observer and black hole. Due to gravitational lensing, light from the source bends around the black hole and makes its way out to the observer, imaging the black hole boundary. The study for Kerr and Kerr-Newman black holes were carried out in \cite{5},\cite{6} and \cite{7}. With improvement in observational data from time to time, black hole shadows were studied in GR \cite{8}-\cite{f}, modified gravity theory \cite{26}-\cite{32} as well as higher dimensional gravity theories \cite{33}-\cite{35}. Besides, black hole shadow analysis were carried out considering expansion of the universe \cite{36},\cite{37}.

\noindent General relativity has impressed us time and again by its impeccable predictions. But it still lacks to incorporate two of the major constituents of the universe- dark energy and dark matter. These two comprise nearly 95-96 \% of the total mass-energy content of the universe. Studies were conducted using dark energy models comprising of cosmological constant ($\Lambda$) \cite{38},\cite{39}, quintessence \cite{40},\cite{41}, etc. But here we are interested in studying  the effects of dark matter ($DM$). Dark matter are non-baryonic and non-luminous in nature comprising about 27 \% of the mass-energy content of the universe. Many observations such as galaxy rotation curves \cite{42} and dynamical motion of galaxy clusters \cite{43} predicted their existence. The widely accepted model for dark matter is the cold dark matter model ($CDM$) whose primary candidate are WIMPs. But the $CDM$ model breaks down at small scales as has been studied in detail in the review work  in \cite{001}. In order to account for the drawbacks of the $CDM$ model, warm and fuzzy dark matter models have been proposed. These models fall into the category of perfect fluid dark matter ($PFDM$). The perfect fluid model was first introduced in \cite{44}, \cite{45} and then further works were carried out in \cite{46}. Recently, $PFDM$ model has been studied using cosmological constant in \cite{47}-\cite{49}. Also shadow of rotating black hole in $PFDM$ has been worked out in \cite{50}. In \cite{1}, circular geodesics were studied in detail in the black hole spacetime with $PFDM$. Black holes surrounded by $PFDM$ in Rastall gravity was studied in \cite{x}. Evidence on the presence of dark matter near black holes have been discussed in \cite{68}. Besides there is a review work on the analytical study of black hole shadow in \cite{1aa}. 

\noindent We plan to study the shadow of charged rotating black hole in perfect fluid dark matter ($PFDM$) surrounded by plasma. In a general scenario  black holes remain surrounded by material media. The recent observations of $M87^{*}$ central black hole supports the presence of plasma \cite{67}. Hence we take interest to investigate the general case where black holes are surrounded by some medium. Study of black hole shadows considering plasma has been conducted in \cite{51}-\cite{60}. We aim to study black hole shadow considering radial power law distribution of plasma. Also we wish to consider cases of both homogeneous \cite{61} and inhomogeneous plasma. Besides, we wish to consider plasma whose frequency ($\omega_p$) depends on both $r$ and $\theta$ coordinates.

\noindent The paper is organised as follows. In section \ref{sec1}, we give a brief overview of the literature in studies of black hole shadow in general and in presence of plasma. In section \ref{sec2}, we discuss the system of rotating charged black hole in $PFDM$. Then in section \ref{sec3}, we continue the subsequent discussion of the black hole system in presence of plasma. Later in section \ref{sec4}, we discuss the circular geodesics. Here we focus in studying the co-rotating and counter rotating photon orbits moving in the equatorial plane. Then we  study the black hole shadow in both arbitrary and near equatorial plane. In section \ref{sec5}, we study the impact of the spacetime parameters \Big($a$, $Q$, $\chi$, $k$, $\frac{\omega_c}{\omega_0}$\Big) on the shadow of the black hole. Later in section \ref{sec6} we look into the effective potential ($V_{eff}$) as encountered by the photons in the black hole spacetime. Further we compute the shadow radius $R_s$, angular shadow radius $\theta_d$ in section \ref{observation} and constraint various parameters by using the observed value of $\theta_d$, obtained from $M87^*$ data. Finally in section \ref{sec7}, we conclude by discussing our observations in detail. We work in geometric units, where we consider $c = G = 1$. Apart from the discussion and analysis, all our mathematical calculations and results are obtained using $M=1$.

\section{Black hole spacetime in perfect fluid dark matter}\label{sec2}
We consider a charged black hole surrounded by perfect fluid dark matter ($PFDM$). The corresponding action  in (3+1) dimensions is given as \cite{1}
\begin{equation}\label{x1}
 	S=\int d^4 x\sqrt{-g}\Big(\frac{R}{16\pi}- \frac{1}{4}F^{\mu \nu}F_{\mu \nu} +\mathcal{L}_{DM}\Big)  
\end{equation}
with $R$ being the Ricci scalar and $F_{\mu \nu}$ being the Maxwell field strength tensor. $\mathcal{L_{DM}}$ gives the Lagrangian density of the $PFDM$. Extremizing the action gives the Einstein field equations \cite{48} 
\begin{equation}\label{x2}
	R_{\mu \nu}-\frac{1}{2}g_{\mu \nu}R=8\pi \Big( T_{\mu\nu}^{M}-T_{\mu\nu}^{DM}\Big)~.
\end{equation}
 In the above equation, $T_{\mu\nu}^{M}$  and $T_{\mu\nu}^{DM}$ are the  energy-momentum tensor of ordinary matter and $PFDM$ respectively. The components of energy-momentum tensor in both cases can be expressed as  \cite{47}, \cite{x}
\begin{eqnarray}
(T^{\mu}_{\;\;\nu})^{M} = diag\left(-\frac{Q^2}{8\pi  r^4}, -\frac{Q^2}{8\pi  r^4}, \frac{Q^2}{8\pi r^4}, \frac{Q^2}{8\pi  r^4}\right)
\end{eqnarray}
\begin{eqnarray}
(T^{\mu}_{\;\;\nu})^{DM} = diag(-\rho, p_r, p, p)
\end{eqnarray}
where $Q$ is the black hole charge, $\rho$ is the energy density and  $p_r$, $p$ and $p$ are pressures of $PFDM$ in the three directions respectively. In order to solve the Einstein field equations, we assume a static, spherically symmetric metric ansatz of the form
\begin{equation}
	ds^2 =-e^{2\gamma}dt^2 + e^{-2\gamma}dr^2 + r^2 (d\theta ^2 + \sin ^2 \theta d\phi^2)
\end{equation}
where $\gamma$ is a function of $r$ only. Replacing the metric ansatz in eq.\eqref{x2}, the different components of Einstein equations take the form
\begin{equation}\label{6a}
	e^{2\gamma}\Big(\frac{1}{r^2}+\frac{2\gamma^{'}}{r}\Big)-\frac{1}{r^2}=8\pi  \Big(-\frac{Q^2}{8\pi  r^4} +\rho \Big)
\end{equation}
\begin{equation}\label{7a}
	e^{2\gamma}\Big(\frac{1}{r^2}+\frac{2\gamma^{'}}{r}\Big)-\frac{1}{r^2}=8\pi  \Big(-\frac{Q^2}{8\pi  r^4} - p_r\Big)
\end{equation}	
\begin{equation}\label{8a}
	e^{2\gamma}\Big(\gamma{''}+2\gamma{'}^2+\frac{2\gamma^{'}}{r}\Big)=8\pi  \Big(\frac{Q^2}{8\pi  r^4} - p\Big)
\end{equation}
where prime ($'$) denotes derivative with respect to the radial coordinate ($r$). Subtracting eq.\eqref{7a} from eq.\eqref{6a} gives
\begin{equation}
    \rho + p_r = 0~~ \Rightarrow ~~  p_r = -\rho~. 
\end{equation}
Taking the equation of state for the $PFDM$ as $\frac{p}{\rho}=(\delta - 1)$, and taking the ratio of the above eq.(s) \eqref{6a} and \eqref{8a}, we get
\begin{equation}\label{1203}
		e^{2\gamma}\Big(\gamma{''}+2\gamma{'}^2+\frac{2\gamma^{'}}{r}\Big)-\frac{Q^2}{ r^4}=(1-\delta)\Bigg[e^{2\gamma}\Big(\frac{1}{r^2}+\frac{2\gamma^{'}}{r}\Big)-\frac{1}{r^2}+\frac{Q^2}{ r^4}\Bigg]~.
\end{equation}
To solve the above equation, we set $2\gamma=\ln(1-U)$, which yields
\begin{equation}\label{1204}
 U^{''}+ 2\delta\frac{U^{'}}{r} + 2(\delta -1)\frac{U}{r^2} + 2(2-\delta)\frac{Q^2}{r^4}=0~.
\end{equation}
Eq.\eqref{1204} can be solved for different values of $\delta$ \cite{46}. In this case, we are particularly interested in the solution for $\delta=\frac{3}{2}$ \cite{44,46}. For $\delta=\frac{3}{2}$, eq.(\ref{1204}) reduces to the following form
\begin{equation}
	r^2 U^{''}+3rU^{'} + U + \frac{Q^2}{r^2}=0~.
\end{equation}
The solution of the above equation is obtained to be
\begin{equation}
	U(r)=\frac{r_s}{r}-\frac{Q^2}{r^2}-\frac{\chi}{r}ln\Big(\frac{r}{|\chi|}\Big)
\end{equation}
where $r_s$ and $\chi$ are integration constants. In order to evaluate $r_s$, we set $Q=0$ and $\alpha=0$. In this limit, weak field approximation gives $r_s = 2M$. Thus, the lapse function takes the form
\begin{eqnarray}\label{05}
	 e^{2\gamma} = 1-U = 1-\frac{2M}{r}+\frac{Q^2}{r^2}+\frac{\chi}{r}ln\Big(\frac{r}{|\chi|}\Big)~.
\end{eqnarray}
The spacetime metric with $e^{2\gamma}=f(r)$ takes the form
\begin{equation}
   	ds^2 =-f(r)dt^2 + \frac{1}{f(r)}dr^2 + r^2 (d\theta ^2 + \sin ^2 \theta d\phi^2) 
\end{equation}
with $f(r)$ given by
\begin{equation}
    f(r)=1-\frac{2M}{r}+\frac{Q^2}{r^2}+\frac{\chi}{r}ln\Big(\frac{r}{|\chi|}\Big)~.
\end{equation}
Replacing the above solution of eq. \eqref{05} in eq.\eqref{6a}, we get 
\begin{equation}\label{08}
     \rho = \frac{\chi}{8\pi r^3}
\end{equation}
where $\rho$ is the energy density of dark matter and for $c=1$ corresponds to the mass density. Since the energy density must be positive due to the weak energy condition, that is, $\rho \geq 0$, this implies that $\chi \geq 0$. Hence the dark matter parameter $\chi$ in the black hole solution (eq.\eqref{05}) is positive \cite{49,50}. It is important to note that if we incorporate $T_{\mu \nu}^{DM}$ in eq.\eqref{x2} with a positive sign, then the weak energy condition would imply $\chi \leq 0$. This would then result in a completely different black hole solution \cite{1b}. We shall not investigate this solution here. We point out that the bound on $\chi$ is $0 < \chi < 2M$ and $-7.18M < \chi < 0$ in the latter case \cite{47}. It is also important to note from eq.\eqref{08} that with increase in the dark matter density $\rho$, the dark matter parameter $\chi$ increases. The rotating version of the solution was obtained using Newman-Janis algorithm in \cite{1}. The solution has the form

\begin{align}\label{7}
ds^2 =-\frac{1}{\zeta^2}\Big(\Delta -a^2 \sin^2 \theta\Big)dt^2 +\frac{\zeta^2}{\Delta}dr^2 + \zeta^2 d\theta^2 -\frac{2a\sin^2 \theta}{\zeta^2}\Big[r^2 + a^2 - \Delta \Big] dtd\phi \nonumber \\
	+ \sin^2 \theta \Big[r^2 + a^2 + \frac{a^2 \sin^2 \theta}{\zeta^2}\Big(r^2 + a^2 - \Delta\Big)\Big]d\phi^2	 
\end{align}\\

\noindent with 
\begin{equation}
\Delta=r^2 +a^2 -2Mr +Q^2+\chi r \ln\Big(\frac{r}{|\chi| }\Big) , ~ \zeta^2 = r^2 + a^2 \cos^2 \theta~.
\end{equation}

\begin{figure}[H]
  \flushleft
  \begin{minipage}[b]{0.45\textwidth}
   {\includegraphics[width=\textwidth]{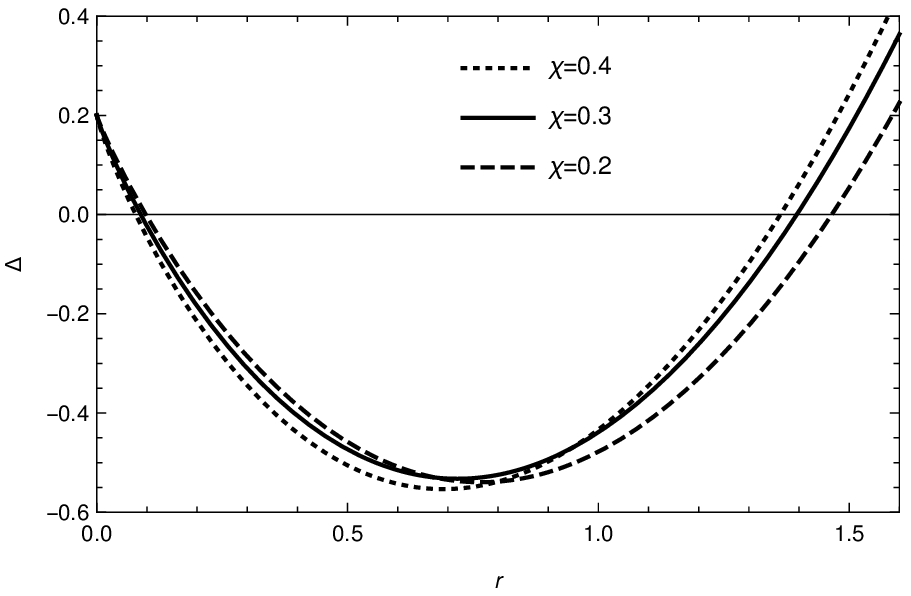}}
    \end{minipage}
\hspace{1.5cm}
 \begin{minipage}[b]{0.45\textwidth}
    {\includegraphics[width=\textwidth]{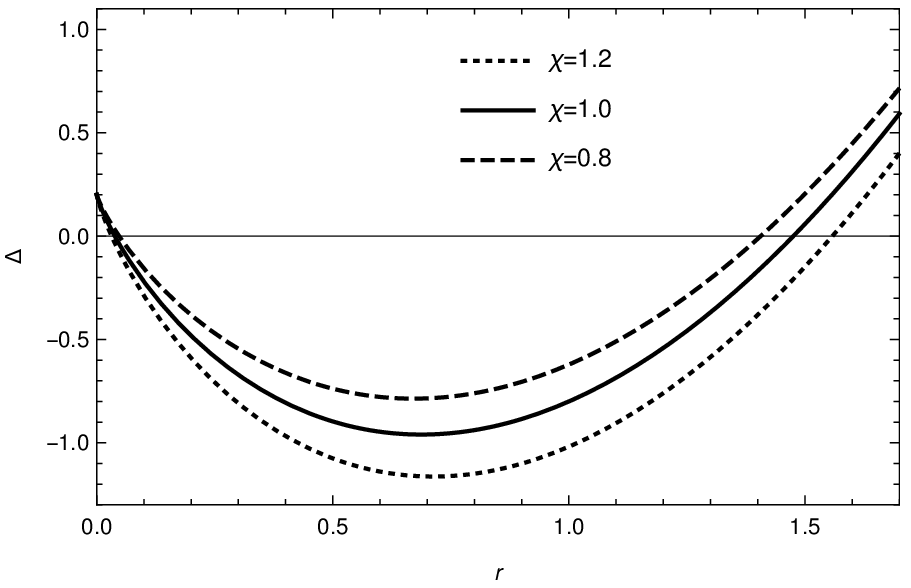}}
     \end{minipage}
  \caption{Plot for $\Delta(r)$ with respect to $r$ for $a$=0.4, $Q$=0.2 for $\chi < \chi_c$(left panel) and  $\chi > \chi_c$(right panel), $\chi_c=0.498$.}
  \label{12}
\end{figure}

\noindent  The black hole event horizon is obtained from the condition $\Delta = 0$. On analysing the spacetime, we find that the black hole has two horizons. One of the horizons is close to $r=0$ but not at $r=0$ which can be seen from the plots. The other horizon is close to $r=1.5$, where we have set $M=1$. The presence of $PFDM$ does not increase the number of horizons of the black hole. Yet it does modify the location of the horizon surface which can be observed by analysing $\Delta$. 

\begin{table}[h]\label{tablea}
\centering
 	\begin{tabular}{|c|c|c|} 	 	
 		\multicolumn{3}{c}{}\\
 		\hline
    	$a$ & $Q$ & $\chi_c$ \\
 		\hline
 		0.1 & 0.2 & 0.529\\ 
 		\hline
 		0.3 & 0.2 &  0.513\\
 		\hline
 		0.5 & 0.2 & 0.477\\ 
 		\hline
 	\end{tabular}
 	\hspace{2.5cm}
 	\begin{tabular}{|c|c|c|} 	 	
 		\multicolumn{3}{c}{}\\
 		\hline
 		$a$ & $Q$ & $\chi_c$ \\
 		\hline
 		0.4 & 0.0 & 0.508\\ 
 		\hline
 		0.4 & 0.3 &  0.488\\
 		\hline
 		0.4 & 0.6 & 0.414\\ 
 		\hline
 	 	\end{tabular}
 	\caption{\footnotesize Table showing the variation of the critical value of $PFDM$ parameter $\chi_c$ with black hole spin $a$ and charge $Q$.}
 	\label{Fig11}
 \end{table} 
\noindent We observe from Fig.\ref{Fig11} that for fixed values of  $M$, $a$ and $Q$, as we increase the $PFDM$ parameter $\chi$, the outer event horizon ($r_{h+}$) initially decreases upto a critical value ($\chi_c$) and then starts increasing. The inner event horizon ($r_{h-}$) decreases all through. The reason for such an observation can be assigned to the fact that $PFDM$ contributes to the mass of the system \cite{50}. We can explain this observation by asserting that the system is a composition of two parts, original black hole of mass $M$ and the black hole corresponding to $PFDM$ with mass $M_{0}$. When $\chi < \chi_c$, then the total system mass is given by $M$ with an inhibition coming from the $PFDM$ mass $M_{0}$. As $\chi$ increases, the inhibition increases which gets reflected in the black hole event horizon ($r_{h+}$). Just at the point when $\chi \geq \chi_c$, $PFDM$ mass $M_{0}$ becomes greater than the original black hole mass $M$ and hence the effective mass and the effective horizon is depicted by the mass of the $PFDM$. The reason for this is that the dark matter density is related directly to the dark matter parameter $\chi$ (eq.\eqref{08}). Thus, for $\chi > \chi_c$, increase in $\chi$, results in increase of the event horizon radius ($r_{h+}$). 
\section{Rotating charged black hole with perfect fluid dark matter immersed in plasma} \label{sec3}
Here we consider the rotating charged black hole surrounded by $PFDM$ immersed in plasma. We assume that there is no interaction between $PFDM$ and plasma. The consideration of plasma is realistic, since most black holes are surrounded by material media. Plasma is a dispersive medium where light rays deviate depending on their frequency. The Hamiltonian for the light rays can be derived using Maxwell's equations considering the source to be made of two charged fluids (electrons and ions). For the study of plasma in curved background, we can consider magnetised \cite{62} as well as non-magnetised plasma \cite{63}. In our study we consider non-magnetised, pressureless (dustlike) plasma. The Hamiltonian for light rays in presence of plasma in curved spacetime is provided in \cite{64}. It is applicable even in our case, since dark matter doesn't interact with plasma by other means except gravity. The Hamiltonian has the form \cite{51}
\begin{equation}\label{222}
    H(x^{\mu}, p_{\mu}) = \frac{1}{2}\Big[ g^{\mu \nu}p_{\mu}p_{\nu} + \omega_p (r) ^2 \Big] 
\end{equation}
where $\omega_p(r)$ is the plasma frequency considered to have radial dependence only.
This is a simplified assumption for a rotating black hole since the plasma frequency should then be a function of both $r$ and $\theta$. The refractive index ($n$) of the material medium depends on both the plasma frequency ($\omega_p$) as well the frequency of photons ($\omega$) measured by a static observer. The expression of $n (r,\omega)$ has the form \cite{65}
\begin{equation}
    n^2(r, \omega) = 1 - \Big(\frac{\omega_p(r)}{\omega}\Big)^2~.
\end{equation}
For an observer having 4-velocity $u^\mu$, the effective energy of photon as measured by the observer is $\hbar \omega = -p_\mu u^\mu$. In our case, the observer is static, hence $\hbar\omega = -p_{0} u^0 =-p_0 \sqrt{-g^{00}}$ \cite{65}. Replacing the expressions for the refractive index of plasma medium and photon energy in eq.\eqref{222}, the Hamiltonian becomes
\begin{equation}\label{223}
    H(x^{\mu}, p_{\mu}) = \frac{1}{2}\Big[ g^{\mu \nu}p_{\mu}p_{\nu} - (n^2 -1)\Big(p_0 \sqrt{-g^{00}}\Big)^2 \Big]~. 
\end{equation}
In order to solve for the geodesics, we need to have an explicit form for the plasma frequency $\omega_p$. Since $\omega_p$ depends only on $r$, plasma can have many forms of distribution. Here we assume an extensively studied distribution in the literature, the radial power law distribution, with $\omega_p$ given as \cite{65}
\begin{equation}\label{2x}
   \omega_p ^2 = \frac{4\pi e^2 N(r)}{m_e} 
\end{equation}
where $e$ is the electronic charge, $m_e$ is the mass of the electron and $N(r)$ being the plasma number density. Note that eq.\eqref{2x} is
relevant when the spacetime is spherically symmetric. 

\noindent The plasma density following \cite{64}, \cite{65} has the form
\begin{equation}
    N(r)=\frac{N_0}{r^h}~.
\end{equation}
where $N_0$ is a constant, $h$ can take integer values with $h \geq 0$. The final form of the refractive index $n(r)$ along with $\frac{\omega_p}{\omega}$ becomes
\begin{equation}
  \Big(\frac{\omega_p}{\omega}\Big)^2 = \frac{k}{r^h}~~;~~n(r)=\sqrt{1-\frac{k}{r^h}}~.  
\end{equation}
Here $k$ is a constant which gives the weightage of plasma around the black hole, also $h$ = 1,2,3 \cite{65}. In this study we shall consider $h=0$ which results in $n(r)=$constant corresponding to homogeneous plasma media. Later we will consider the simplest dependence on $r$ with $h=1$. The second case will be regarded as inhomogeneous plasma distribution.

\noindent Also we consider a case where the plasma frequency $\omega_p (r, \theta)$ is dependent on both $r$ and $\theta$ such that
\begin{equation}\label{200}
    \omega_p (r, \theta)^2 = \frac{f_r (r) + f_{\theta} (\theta)}{r^2 + a^2 \cos^2 \theta}~.
\end{equation}
We have carried out the shadow analysis considering this general case in the subsequent 
discussion. 	Note that the above form of the plasma distribution (eq.\eqref{200}) is relevant when the spacetime is axially-symmetric.

\section{Study of circular geodesics in plasma medium}\label{sec4}
\subsection{Co-rotating and counter rotating photon orbits}
The photons in the rotating black hole spacetime can move along the direction of black hole spin (co-rotating photons) as well as opposite to that of black hole spin (counter rotating photons). In order to determine the radius ($r_p$) of those photon orbits, we need to determine the radial geodesic equation and impose the condition for circular geodesics. For this, we need to use the Hamiltonian ($\mathcal{H}$) to determine the trajectories of the particles in the equatorial plane. The Hamiltonian given in eq.\eqref{223} takes the form
\begin{equation}\label{20}
    \mathcal{H}=\frac{1}{2}\Big[ g^{\mu \nu}p_{\mu}p_{\nu} + (n^2 -1)g^{00}p_0 ^2 \Big]
\end{equation}
with $\mathcal{H}=0$ for photons. Since we are interested in the geodesics of the equatorial plane, we set $\theta = \frac{\pi}{2}$ and thus $\dot{\theta}=0$. \noindent The geodesics can be calculated using Hamilton's equation of motion given by 
\begin{equation}
  \dot{x}^{\mu}=\frac{\partial \mathcal{H}}{\partial p_{\mu}}~~;~~  \dot{p}_{\mu}=-\frac{\partial \mathcal{H}}{\partial x^{\mu}}~. 
\end{equation}
The Hamiltonian depends on the metric ($g_{\mu \nu}$) as well as on the refractive index of plasma ($n(r)$). The metric has no explicit dependence on $x_0 (=t)$ and $x_3 (=\phi)$. So the Hamilton's second equation of motion  gives two constants of motion $p_0 = -E$  and $p_3 = L_{\phi}$. $E$ and $L_{\phi}$ respectively give the energy per unit mass and angular momentum per unit mass of the photon as observed by a stationary observer at infinity. The geodesics corresponding to $t$ and $\phi$ in the equatorial plane becomes

\begin{equation}
    r^2 \dot{t} = \frac{r^2 + a^2}{\Delta}\Big[n^2(r^2 + a^2)E - aL_{\phi}\Big] + a\Big[L_{\phi} - an^2 E \Big]
\end{equation}
\begin{equation}
    r^2 \dot{\phi} = \frac{a}{\Delta}\Big[(r^2 + a^2)E - aL_{\phi}\Big] + \Big[L_{\phi} - a E \Big]~.
\end{equation}
Now  $p_r = \frac{\partial \mathcal{L}}{\partial \dot{r}} = \frac{\partial S}{\partial r} = g_{rr} \dot{r}^2$ with $\mathcal{L}=\frac{1}{2}g_{\mu \nu}\dot{x}^{\mu}\dot{x}^{\nu}$. The dot in the above equations correspond to derivative with respect to the affine parameter $\lambda$. Since light rays cannot be parametrized in terms of proper time ($\tau$), so we need to parametrize them with some other parameter. This parametrization is done using the affine parameter ($\lambda$). With the above equations in hand, we get from eq.\eqref{20}
\begin{equation}\label{21}
    \dot{r}^2 = \frac{1}{r^4}\Bigg[\Big(E(r^2 + a^2) - aL_{\phi}\Big)^2 - \Delta(aE-L_{\phi})^2 + (n^2 - 1)\Big(E^2 (r^2 + a^2)^2 -\Delta E^2 a^2 \Big) \Bigg]~.
\end{equation}
In order to obtain the two kinds of orbits discussed above, we define the impact parameter as $\frac{L_{\phi}}{E}=D$, in terms of which the radial eq.\eqref{21} looks as
\begin{equation}\label{22}
    \dot{r}^2 = \frac{E^2}{r^4}\Bigg[\Big((r^2 + a^2) - aD\Big)^2 - \Delta(a-D)^2 + (n^2 - 1)\Big( (r^2 + a^2)^2 -\Delta a^2 \Big) \Bigg]~.
\end{equation}

\noindent Rearranging eq.\eqref{22}, we have
\begin{equation}\label{23}
   \frac{r^2\dot{r}^2}{E^2} =\frac{1}{r^2}\Bigg[\Big((r^2 + a^2) - aD\Big)^2 - \Delta(a-D)^2 + (n^2 - 1)\Big( (r^2 + a^2)^2 -\Delta a^2 \Big) \Bigg]=F(r)~.
\end{equation}
Imposing the condition for circular geodesics, that is, $F(r)=0=F'(r)$, we get
\begin{equation}\label{47}
   r^2 + (a^2 - D^2) + \frac{2M}{r}(a-D)^2 -\frac{Q^2}{r^2}(a-D)^2 - \frac{\chi}{r}(a-D)^2 \ln\Big(\frac{r}{|\chi|}\Big) + 
   (n^2 -1)\Big(r^2 +a^2 +a^2\Big(\frac{2M}{r} - \frac{Q^2}{r^2} - \frac{\chi}{r}\ln\Big(\frac{r}{|\chi|}\Big)\Big)\Big)=0~. 
\end{equation}
\begin{eqnarray}\label{24}
2r -\frac{2M}{r^2}(a-D)^2 + \frac{2Q^2}{r^3}(a-D)^2 + \frac{\chi}{r^2}(a-D)^2\ln(\frac{r}{|\chi|})-\frac{\chi}{r^2}(a-D)^2 + \nonumber \\ (n^2 -1)\Big(2r-a^2\Big(\frac{2M}{r^2} - \frac{2Q^2}{r^3} - \frac{\chi}{r^2}\ln(\frac{r}{|\chi|}) + \frac{\chi}{r^2} \Big)\Big) + 2nn' \Big(r^2 +a^2 +a^2\Big(\frac{2M}{r} - \frac{Q^2}{r^2} - \frac{\chi}{r}\ln\Big(\frac{r}{|\chi|}\Big)\Big)\Big)=0
\end{eqnarray}
where, $n^{'}\equiv \frac{dn}{dr}$.
Solving for $(a-D)$ in eq.\eqref{24}, we have
\begin{footnotesize}
\begin{eqnarray}
(a-D)=\pm \sqrt{\frac{2r^5 + (n^2 -1)\Big[2r^5 - a^2\Big(2Mr^2 - 2Q^2 r - \chi r^2 \ln(\frac{r}{|\chi|})+\chi r^2\Big)\Big] + 2nn'\Big[r^4(r^2 + a^2) + a^2\Big(2Mr^3 - Q^2 r^2 - \chi r^3 \ln(\frac{r}{|\chi|})\Big)\Big]}{2Mr^2 - 2Q^2 r - \chi r^2 \ln(\frac{r}{|\chi|})+\chi r^2}}~.
\end{eqnarray}
\end{footnotesize}
Thus the impact parameter becomes
\begin{footnotesize}
\begin{eqnarray}\label{50}
D=a\mp \sqrt{\frac{2r^5 + (n^2 -1)\Big[2r^5 - a^2\Big(2Mr^2 - 2Q^2 r - \chi r^2 \ln(\frac{r}{|\chi|})+\chi r^2\Big)\Big] + 2nn'\Big[r^4(r^2 + a^2) + a^2\Big(2Mr^3 - Q^2 r^2 - \chi r^3 \ln(\frac{r}{|\chi|})\Big)\Big]}{2Mr^2 - 2Q^2 r - \chi r^2 \ln(\frac{r}{|\chi|})+\chi r^2}}~.
\end{eqnarray}
\end{footnotesize}
The $\mp$ sign corresponds to counter and co-rotating geodesics of photons moving in the black hole spacetime. In case of background devoid of plasma, that is, $n=1, n' =0$ we have 
\begin{equation}
    D=a\mp \sqrt{\frac{2r^5}{2Mr^2 - 2Q^2 r - \chi r^2 \ln(\frac{r}{|\chi|})+\chi r^2}}
\end{equation}
which corresponds to the expression derived in \cite{1}.
Replacing $D$ from eq.\eqref{50} in eq.\eqref{47} we get an equation in $r$ devoid of the constants $E$ and $L_{\phi}$ which depends only on the spacetime parameters ($M$, $Q$, $\chi$, $n$). The solution of the equation gives the photon orbit radius ($r_p$) both for co-rotating and counter rotating orbits. The Table(s) below show the photon orbit radius ($r_p$) both for co and counter rotating orbits with variation in plasma parameter ($k$). We obtain the Table(s) below with $M=1$.
\begin{table}[h]\label{table}
\centering
 	\begin{tabular}{|c|c|} 	 	
 		\multicolumn{2}{c}{\textbf{ $\chi$ =0.2}}\\
 		\hline
 		$k$ & $r_{p1}$ \\
 		\hline
 		0.0 & 1.645\\ 
 		\hline
 		0.2 & 1.480  \\
 		\hline
 		0.27 & 1.381\\ 
 		\hline
 	\end{tabular}
 	\hspace{1.5cm}
 	\begin{tabular}{|c|c|} 	 	
 		\multicolumn{2}{c}{\textbf{ $\chi$ =1.0}}\\
 		\hline
 		$k$ & $r_{p1}$ \\
 		\hline
 		0.0 & 1.742\\ 
 		\hline
 		0.2 &1.632   \\
 		\hline
 		0.38 & 1.418\\ 
 		\hline
 	\end{tabular}
 	\hspace{1.5cm}
 	\begin{tabular}{|c|c|} 	 	
 		\multicolumn{2}{c}{\textbf{ $\chi$ =0.2}}\\
 		\hline
 		$k$ & $r_{p1}$ \\
 		\hline
 		0.0 & 1.645\\ 
 		\hline
 		0.2 & 1.507    \\
 		\hline
 		0.31 & 1.385\\ 
 		\hline
 	\end{tabular}
 	\hspace{1.5cm}
 	\begin{tabular}{|c|c|} 	 	
 		\multicolumn{2}{c}{\textbf{ $\chi$ =1.0}}\\
 		\hline
 		$k$ & $r_{p1}$ \\
 		\hline
 		0.0 & 1.742\\ 
 		\hline
 		0.2 & 1.632    \\
 		\hline
 		0.4 & 1.492\\ 
 		\hline
 	\end{tabular}
 	\caption{\footnotesize Radius ($r_{p1}$) for co-rotating (prograde) photon orbits with variation in plasma parameter ($k$) with black hole spin $a$=0.5 and charge $Q$=0.3. The first two are for homogeneous plasma \Big($n=\sqrt{1-k}$\Big) and the next two for inhomogeneous plasma \Big($n=\sqrt{1-\frac{k}{r}}$\Big). }
 	\label{Fig1}
 \end{table} 
 
 \noindent Table \ref{Fig1} show that with increase in the plasma parameter $k$, the radius of co-rotating photon orbits $r_{p1}$ decrease in case of both homogeneous and inhomogeneous plasma. Besides they exist closer to the event horizon ($r_{h+}$) of the black hole. The effect of plasma parameter $k$ on the radius of the orbits remains same for both $\chi < \chi_c$ and $\chi > \chi_c$. The critical value of the $PFDM$ parameter for black hole spin $a=0.5$ and charge $Q=0.3$ is 0.467. Also we find that the photon orbits do not exist for all possible values of the plasma parameter $k$. After a certain critical value of plasma parameter $k=k_c$, the photon radius drops below the event horizon radius ($r_{h+}$) and thus has no physical existence. Thus there is a bound to the value of $k$ depending on the combinations of the black hole parameters ($M$, $Q$, $\chi$).
\begin{table}[h]\label{table}
\centering
 	\begin{tabular}{|c|c|} 	 	
 		\multicolumn{2}{c}{\textbf{ $\chi$ =0.2}}\\
 		\hline
 		$k$ & $r_{p2}$ \\
 		\hline
 		0.0 & 2.782\\ 
 		\hline
 		0.2 &2.825    \\
 		\hline
 		0.4 & 2.886\\ 
 		\hline
 	\end{tabular}
 	\hspace{1.5cm}
 	\begin{tabular}{|c|c|} 	 	
 		\multicolumn{2}{c}{\textbf{ $\chi$ =1.0}}\\
 		\hline
 		$k$ & $r_{p2}$ \\
 		\hline
 		0.0 & 2.608\\ 
 		\hline
 		0.2 &2.640    \\
 		\hline
 		0.4 & 2.685\\ 
 		\hline
 	\end{tabular}
 	\hspace{1.5cm}
 	\begin{tabular}{|c|c|} 	 	
 		\multicolumn{2}{c}{\textbf{ $\chi$ =0.2}}\\
 		\hline
 		$k$ & $r_{p2}$ \\
 		\hline
 		0.0 & 2.782\\ 
 		\hline
 		0.2 &2.761    \\
 		\hline
 		0.4 & 2.737\\ 
 		\hline
 	\end{tabular}
 	\hspace{1.5cm}
 	\begin{tabular}{|c|c|} 	 	
 		\multicolumn{2}{c}{\textbf{ $\chi$ =1.0}}\\
 		\hline
 		$k$ & $r_{p2}$ \\
 		\hline
 		0.0 & 2.608\\ 
 		\hline
 		0.2 &2.587    \\
 		\hline
 		0.4 & 2.564\\ 
 		\hline
 	\end{tabular}
 	\caption{\footnotesize Radius ($r_{p2}$) for counter rotating (retrograde) photon orbits with variation in plasma parameter ($k$) with black hole spin $a$=0.5 and charge $Q$=0.3. The first two are for homogeneous plasma \Big($n=\sqrt{1-k}$\Big) and the next two for inhomogeneous plasma \Big($n=\sqrt{1-\frac{k}{r}}$\Big). }
 	\label{Fig2}
 \end{table} 
 
\noindent Table \ref{Fig2} show that with increase in plasma parameter $k$, the radius of counter rotating photon orbits $r_{p2}$ increase in case of homogenous plasma distribution whereas it decreases in case of inhomogenous distribution. The effect of plasma parameter $k$ on the radius of the orbits remains same for both $\chi < \chi_c$ and $\chi > \chi_c$. Here too only for a certain combination of spacetime parameters and plasma parameter $k$ does the photon orbits exist which results in a bound on the plasma parameter $k$.

\subsection{Black hole shadow}
In this section, we wish to calculate the black hole shadow. Shadows are formed due to bending of light rays near regions of strong gravity. In case of black holes, when light from a distant source comes near to it, light rays get deflected. The deflected rays after encircling the black hole either plunges into the black hole or escapes to infinity. These light rays from unstable circular orbits reaches the observer at infinity and creates \textit{circular} or \textit{deformed circular} boundary curve. The dark disk inside the curve is called the black hole shadow. The shadow is formed in the celestial plane characterised by the celestial coordinates $\alpha$ and $\beta$. They are defined for an observer at infinity as
\begin{multicols}{2}
\begin{figure}[H]
  \centering
  \begin{minipage}[b]{0.4\textwidth}
   {\includegraphics[width=\textwidth]{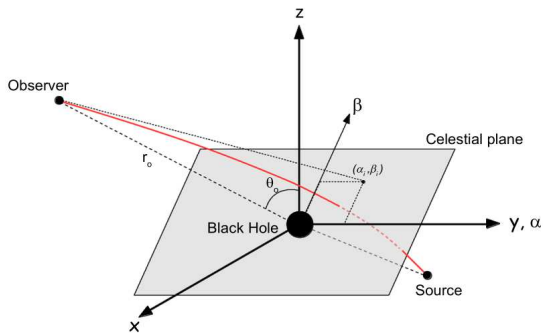}}
    \end{minipage}
    \caption{Celestial coordinates ($\alpha$, $\beta$)}\cite{c}.
  \label{31}
    \end{figure}
\begin{equation}
    \alpha = \lim_{r_0\to \infty} -r_0 ^2 \sin \theta_0 \Big(\frac{d \phi}{d r}\Big)\Bigg|_{r_0, \theta_0}
    \end{equation}
    \begin{equation}
    \beta = \lim_{r_0 \to \infty} r_0 ^2 \Big(\frac{d \theta}{d r}\Big)\Bigg|_{r_0, \theta_0}
 \end{equation}
\end{multicols}

\noindent where the position of the observer is given by the coordinates $r_0$ and $\theta_0$. As can be seen from Fig.\ref{31}, we consider a black hole residing between the source and the observer. We have shown a tangent on the light ray at the point where light reaches the observer. The light ray meets the celestial plane at the point ($\alpha, \beta$). Considering all light rays from all possible directions and drawing tangents that intersect the plane, we get a circular or near circular or deformed circular boundary curve. The celestial coordinate $\alpha$ gives the apparent perpendicular distance of the shadow boundary from the axis of black hole rotation and the coordinate $\beta$ gives the apparent perpendicular distance of the boundary of the shadow from its projection in the equatorial plane. $r_0$ gives the radial distance of the observer from the black hole and $\theta_0$ corresponds to the inclination angle of the observer's line of sight with the axis of rotation.

\subsubsection{Case I}\label{I}
Here we consider the general case by considering the plasma frequency to be a function of both $r$ and $\theta$, that is, $n (r, \theta)$ \cite{1aa},\cite{54},\cite{60}. We start from the Hamiltonian 
\begin{equation}
    \mathcal{H}(x^{\mu}, p_{\mu})=\frac{1}{2}\Big[g^{\mu \nu}p_{\mu}p_{\nu} + \omega_p(r,\theta) ^2\Big]
\end{equation}
with the refractive index defined as 
\begin{equation}
    n(r,\theta)^2 = 1 - \Big(\frac{\omega_p(r,\theta)}{\omega}\Big)^2~.
\end{equation}
Here, $\omega_p (r, \theta)$ gives the plasma frequency and $\omega$ gives the photon frequency as measured by any arbitrary observer in the domain of outer communication, that is between outer event horizon ($r_{h+}$) to infinity. The photon frequency as measured by a stationary observer at infinity is $\omega_0$ which is related to $\omega$ by the relation
\begin{equation}\label{79}
    \omega = \frac{\omega_0}{\sqrt{-g_{00}}}=\frac{\omega_0 \zeta}{\sqrt{\Delta -a^2 \sin^2 \theta}}~.
\end{equation}
We wish to calculate the geodesics by Hamilton's equation of motion. We found that the Hamiltonian ($\mathcal{H}$) is independent of $t$ and $\phi$. Hence the corresponding constants of motion are $p_0 = -E$ and $p_3 = L_{\phi}$ where, $E$ and $L_{\phi}$ correspond to the energy and angular momentum of photons as measured by a stationary observer at infinity. Also $E$ and $\omega_0$ are related as $E=\hbar \omega_0$, which results into $E=\omega_0$ for $\hbar =1$. By using Hamilton's equation of motion $\dot{x}^{\mu}=\frac{\partial H}{\partial p_{\mu}}$, we get the equations for $t$ and $\phi$ as
\begin{equation}
    \zeta^2 \dot{t} = \frac{r^2 + a^2}{\Delta}\Big[(r^2 + a^2)E - aL_{\phi}\Big] + a\Big[L_{\phi} - a E \sin ^2 \theta\Big]
\end{equation}
\begin{equation}\label{44}
    \zeta^2 \dot{\phi} = \frac{a}{\Delta}\Big[(r^2 + a^2)E - aL_{\phi}\Big] + \Big[L_{\phi}\csc^2 \theta - a E \Big]~.
\end{equation}
For evaluating the geodesics of $r$ and $\theta$, we use the Hamilton-Jacobi equation given for photons as
\begin{equation}\label{000}
     \mathcal{H}\Big(x^{\mu}, \frac{\partial S}{\partial x^{\mu}}\Big)=\frac{1}{2}\Big[g^{\mu \nu}\frac{\partial S}{\partial x^{\mu}}\frac{\partial S}{\partial x^{\nu}} + \omega_p(r,\theta) ^2\Big]=0~.
\end{equation}
In order to solve the above eq.\eqref{000}, we assume an ansatz of the form \cite{66}
\begin{equation}
    S=-Et + L_{\phi}\phi + S_r (r) + S_{\theta} (\theta)
\end{equation}
and replace it in eq.\eqref{000} to get
\begin{equation}\label{010}
     -\frac{1}{\Delta}\Big[(r^2 + a^2)E - aL_{\phi}\Big]^2 + (L_{\phi} \csc \theta - aE\sin \theta )^2 +\Delta \Big(\frac{\partial S_r}{\partial r}\Big)^2 + \Big(\frac{\partial S_\theta}{\partial \theta}\Big)^2 + \omega_p ^2 \zeta^2 = 0~.
\end{equation}
To separate the above equation into two equations of $r$ and $\theta$, we need to assume a certain form of $\omega_p ^2(r, \theta)$ which reads \cite{54}
\begin{equation}\label{85}
    \omega_p ^2 (r, \theta)  = \frac{f_r (r) + f_{\theta} (\theta)}{r^2 + a^2 \cos^2 \theta}
\end{equation}
where $f_r (r)$ and $f_{\theta} (\theta)$ are functions of $r$ and $\theta$ only respectively. This form of $\omega_p ^2 (r, \theta)$ leads to the following form for the refractive index $n(r,\theta)$

\begin{equation}
    n(r,\theta)^2 = 1 - \Big(\frac{f_r (r) + f_{\theta} (\theta)}{(r^2 + a^2 \cos^2 \theta)\omega^2}\Big)~.
\end{equation}

\noindent This gives us 
\begin{equation}
     \Big(\frac{\partial S_\theta}{\partial \theta}\Big)^2 + (L_{\phi} \csc \theta - aE\sin \theta )^2 + f_{\theta} (\theta) = \frac{1}{\Delta}\Big[(r^2 + a^2)E - aL_{\phi}\Big]^2 - \Delta \Big(\frac{\partial S_r}{\partial r}\Big)^2 - f_r (r)= constant= \kappa
\end{equation}
with $\kappa$ being the generalised Carter constant used to separate the $r$ and $\theta$ geodesics. The equations take the form
\begin{equation}\label{k5}
   \zeta^2 \dot{\theta}= \pm \sqrt{\Theta (\theta)}
\end{equation}
\begin{equation}\label{k6}
    \zeta^2 \dot{r}= \pm \sqrt{R (r)}
\end{equation}
with $\Theta(\theta)$ and $R(r)$ taking the form
\begin{equation}
    \Theta (\theta)= \kappa- (L_{\phi} \csc \theta - aE\sin \theta )^2 - f_{\theta} (\theta)
\end{equation}
\begin{equation}
    R(r)= \Big[(r^2 + a^2)E - aL_{\phi}\Big]^2 -\Delta\Big(\kappa + f_{r} (r)\Big)~.
\end{equation}
The shadow is formed by the photons moving in the unstable circular geodesics. The conditions for these geodesics are $R(r) = \frac{\partial R(r)}{\partial r} = 0.$ Utilizing these conditions, we get the following equations 
\begin{equation}\label{k1}
    \Big[(r^2 + a^2) - a\xi\Big]^2 = \Delta\Big[\eta + \tilde{f}_r (r)\Big]
\end{equation}
\begin{equation}\label{k2}
  4r\Big[(r^2 + a^2) - a\xi\Big] - \Delta \tilde{f}_r (r)  = \Delta^{'}\Big[\eta + \tilde{f}_r (r)\Big]  
\end{equation}
where $\xi = \frac{L_{\phi}}{E}$, $\eta = \frac{\kappa}{E^2}$ and $\tilde{f}_r (r) = \frac{f_r (r)}{E^2}$. Also we define another variable $\tilde{f}_{\theta} (\theta)= \frac{f_{\theta} (\theta)}{E^2}$ which will be necessary later~. Eliminating $\eta$ from eq.(s) \eqref{k1} and \eqref{k2}, we get an equation in $\xi$ as
\begin{equation}
    A \xi^2 + 2 B \xi + C=0
\end{equation}
where
\begin{align}
    A = a^2 \Delta^{'}~~;~~ B = 2ar\Delta - a\Delta^{'}(r^2 + a^2)~;~~\\ \nonumber
    C = \Big[\Delta^{'}(r^2 + a^2)^2 -4r \Delta (r^2 + a^2) + \Delta ^2 \tilde{f}'_r (r) \Big]
\end{align}
with $\tilde{f}'_r (r)$ giving the derivative of $\tilde{f}_r (r)$ with respect to $r$. Solving for $\xi$, we get
\begin{equation}\label{k3}
    \xi = -\frac{B}{A} \pm \sqrt{\Big(\frac{B}{A}\Big)^2 - \frac{C}{A}}~.
\end{equation}
The shadow can be obtained by considering the negative sign in the above expression. The expression for $\eta$ takes the form
\begin{equation}
    \eta = \frac{1}{\Delta}\Bigg[ (r^2 + a^2) - a\xi \Bigg]^2 - \tilde{f}'_r (r) ~.
\end{equation}
From the expressions of the celestial coordinates $\alpha$ and $\beta$, we observe that for determining the black hole shadow, we need the geodesics for $\phi$, $\theta$ and $r$ as given in eq.(s) \eqref{44}, \eqref{k5} and \eqref{k6} respectively.  The expressions for $\frac{d \phi}{dr}$ and $\frac{d \theta}{dr}$ take the form
\begin{equation}\label{k7}
   \Big(\frac{d \phi}{d r}\Big) = \frac{\frac{a}{\Delta}\Big[(r^2 + a^2)E - aL_{\phi}\Big] + \Big[L_{\phi}\csc^2 \theta - a E \Big]}{\sqrt{\Big[(r^2 + a^2)E - aL_{\phi}\Big]^2 -\Delta\Big(\kappa + f_{r} (r)\Big)}}
\end{equation}
\begin{equation}\label{k8}
   \Big(\frac{d \theta}{d r}\Big) = \sqrt{\frac{\kappa - (L_{\phi} \csc \theta - aE\sin \theta )^2 - f_{\theta} (\theta)}{\Big[(r^2 + a^2)E - aL_{\phi}\Big]^2 -\Delta\Big(\kappa + f_{r} (r)\Big)}}~.
\end{equation}
Using the above eq.(s) \eqref{k7}, \eqref{k8} in the expressions for $\alpha$ and $\beta$, we have
\begin{equation}\label{k9}
    \alpha = -\xi\csc \theta_0~~;~~ \beta =\pm \sqrt{\eta - (\xi \csc \theta_0 - a\sin \theta_0 )^2 - \tilde{f}_{\theta} (\theta_0)}~.
\end{equation}
The shadow can be observed by plotting $\alpha$ and $\beta$ along X and Y axis respectively. We show the plots for two different plasma distribution. One for $f_r (r) =\omega_c ^2 \sqrt{M^3 r}$ and $f_{\theta} (\theta)=0$ \cite{54} and the other one for $f_r (r) =0$ and $f_{\theta} (\theta)= \omega_c ^2 M^2 (1 + 2 \sin^2 \theta)$ \cite{54}. For $f_r (r) =\omega_c ^2 \sqrt{M^3 r}$ and $f_{\theta} (\theta)=0$ (with $M=1$), the expressions in eq.(s) \eqref{k7}, \eqref{k8} and \eqref{k9} take the form

\begin{equation}
   \Big(\frac{d \phi}{d r}\Big) = \frac{\frac{a}{\Delta}\Big[(r^2 + a^2)E - aL_{\phi}\Big] + \Big[L_{\phi}\csc^2 \theta - a E \Big]}{\sqrt{\Big[(r^2 + a^2)E - aL_{\phi}\Big]^2 -\Delta\Big(\kappa + \omega_c ^2 \sqrt{r}\Big)}}
\end{equation}
\begin{equation}
   \Big(\frac{d \theta}{d r}\Big) = \sqrt{\frac{\kappa - (L_{\phi} \csc \theta - aE\sin \theta )^2 }{\Big[(r^2 + a^2)E - aL_{\phi}\Big]^2 -\Delta\Big(\kappa +  \omega_c ^2 \sqrt{r}\Big)}}
\end{equation}
which gives
\begin{equation}
    \alpha = -\xi\csc \theta_0~~;~~ \beta =\pm \sqrt{\eta - (\xi \csc \theta_0 - a\sin \theta_0 )^2}~.
\end{equation}
Also for $f_r (r) =0$ and $f_{\theta} (\theta)= \omega_c ^2 M^2 (1 + 2 \sin^2 \theta)$ (with $M=1$), the expressions in eq.(s) \eqref{k7}, \eqref{k8} and \eqref{k9} take the form

\begin{equation}
   \Big(\frac{d \phi}{d r}\Big) = \frac{\frac{a}{\Delta}\Big[(r^2 + a^2)E - aL_{\phi}\Big] + \Big[L_{\phi}\csc^2 \theta - a E \Big]}{\sqrt{\Big[(r^2 + a^2)E - aL_{\phi}\Big]^2 -\Delta \kappa }}
\end{equation}
\begin{equation}
   \Big(\frac{d \theta}{d r}\Big) = \sqrt{\frac{\kappa - (L_{\phi} \csc \theta - aE\sin \theta )^2 -\omega_c ^2 (1 + 2 \sin^2 \theta) }{\Big[(r^2 + a^2)E - aL_{\phi}\Big]^2 -\Delta \kappa }}
\end{equation}
which gives
\begin{equation}
    \alpha = -\xi\csc \theta_0~~;~~ \beta =\pm \sqrt{\eta - (\xi \csc \theta_0 - a\sin \theta_0 )^2 - \Big(\frac{\omega_c}{\omega_0}\Big) ^2 (1 + 2 \sin^2 \theta_0)}~.
\end{equation}

\noindent The plots are shown in section \ref{sec5} (Figures \ref{10a} and \ref{10b}) and we have set $M=1$.

\subsubsection{Case II}\label{II}
In this section, we plan to investigate the black hole shadow for the case of refractive index $n(r)$ depending only on $r$ and having no $\theta$ dependence. For this we like to evaluate the geodesics of light rays. The Hamiltonian ($\mathcal{H}$) in this case takes the form 
\begin{equation}
    \mathcal{H}=\frac{1}{2}\Big[ g^{\mu \nu}p_{\mu}p_{\nu} + (n^2 -1)g^{00}p_0 ^2 \Big]~.
\end{equation}
\noindent Using Hamilton's equation of motion, we can evaluate the expressions for $t$ and $\phi$ geodesics in an arbitrary plane which reads 
\begin{equation}
    \zeta^2 \dot{t} = \frac{r^2 + a^2}{\Delta}\Big[n^2(r^2 + a^2)E - aL_{\phi}\Big] + a\Big[L_{\phi} - an^2 E \sin ^2 \theta\Big]
\end{equation}
\begin{equation}\label{002}
    \zeta^2 \dot{\phi} = \frac{a}{\Delta}\Big[(r^2 + a^2)E - aL_{\phi}\Big] + \Big[L_{\phi}\csc^2 \theta - a E \Big]~.
\end{equation}
In order to evaluate the geodesics for $r$ and $\theta$, we use the Hamilton-Jacobi equation, which reads 
\begin{equation}\label{11}
    \frac{\partial S}{\partial \lambda}= -\mathcal{H}=-\frac{1}{2}\Big[ g^{\mu \nu}\Big(\frac{\partial S}{\partial x^{\mu}}\Big)\Big(\frac{\partial S}{\partial x^{\nu}}\Big) + (n^2 -1)g^{00}\Big(\frac{\partial S}{\partial x^0}\Big)^2 \Big] ~.
\end{equation}
In order to solve the above equation, we need to choose an ansatz for $S$. Following \cite{66}, we use 
\begin{equation}
S=-Et + L_{\phi}\phi +S_r(r) + S_{\theta}(\theta)~.
\end{equation}
Replacing $S$ in eq.\eqref{11}, we have
\begin{align}\label{19}
    -\frac{(n^2 - 1)}{\Delta}E^2 (r^2 + a^2)^2 -\frac{1}{\Delta}\Big[(r^2 + a^2)E - aL_{\phi}\Big]^2 + (n^2-1) a^2 E^2 \sin^2 \theta -a^2E^2\cos^2 \theta \\ \nonumber
    + L^2 _{\phi}\cot ^2 \theta + (aE - L_{\phi})^2 +\Delta \Big(\frac{\partial S_r}{\partial r}\Big)^2 + \Big(\frac{\partial S_\theta}{\partial \theta}\Big)^2 = 0~.
\end{align}
Now we wish to separate the above equation in $r$ and $\theta$ variables. By inspecting the term $n(r)^2 a^2 E^2 \sin^2 \theta$, we find that it is in general, not possible to separate $n(r)$ from $\sin \theta$. We now investigate two cases.\\

\noindent \textbf{Case IIa}\\
First, we consider $n=n(r)=\sqrt{1-\frac{k}{r}}$. Though it should be noted that for spinning black holes $n=n(r, \theta)$ \cite{1aa,54}, but for simplicity we now consider the $n=n(r)$ only. Now we find that due to such consideration, eq.\eqref{19} is not separable. However, we can separate it into $r$ and $\theta$ equations by considering near equatorial plane as  done in \cite{11,c}. The approximation is taken as $\theta \approx \frac{\Pi}{2} + \epsilon$, where $\epsilon$ is a very small angle. It must be noted that the unstable photon orbits are not restricted to the near equatorial planes, they can travel through any arbitrary plane. But for an observer at infinity, this approximation is valid and gives desirable results for black hole shadow. Besides since we are considering near equatorial plane, the observer should also be placed in the equatorial plane, i.e., $\theta_0 = \frac{\pi}{2}$. This assumption modifies the geodesics which we take into account while evaluating the celestial coordinates ($\alpha, \beta$).

\noindent Choosing a near equatorial plane and setting $\theta = \frac{\Pi}{2} + \epsilon$, we have eq.\eqref{19} as
\begin{align}\label{3}
    -\frac{(n^2 - 1)}{\Delta}E^2 (r^2 + a^2)^2 -\frac{1}{\Delta}\Big[(r^2 + a^2)E - aL_{\phi}\Big]^2 + (n^2 -1)a^2 E^2 \\ \nonumber
     + (aE - L_{\phi})^2 +\Delta \Big(\frac{\partial S_r}{\partial r}\Big)^2 + \Big(\frac{\partial S_\epsilon}{\partial \epsilon}\Big)^2 = 0~.
\end{align}
Introducing Carter constant $\kappa$ \cite{66} as in the earlier case, we can split the equation into two parts as
\begin{align}\label{3}
    \frac{(n^2 - 1)}{\Delta}E^2 (r^2 + a^2)^2 + \frac{1}{\Delta}\Big[(r^2 + a^2)E - aL_{\phi}\Big]^2 - (n^2-1) a^2 E^2 \\ \nonumber
     - (aE - L_{\phi})^2 -\Delta \Big(\frac{\partial S_r}{\partial r}\Big)^2 = \Big(\frac{\partial S_\epsilon}{\partial \epsilon}\Big)^2 = \kappa~.
\end{align}
Finally, using \( \frac{\partial \mathcal{L}}{\partial \dot{x}^ \mu} = \frac{\partial S}{\partial x ^ \mu} \) ,
 where \(\mathcal{L} = \frac{1}{2}g_{\mu \nu}\dot{x}^ \mu \dot{x}^ \nu \) , we get equation(s) for $r$ and $\epsilon$ as
\begin{equation}
     r^2 \dot{r} = \sqrt{R(r)}
\end{equation}
\begin{equation}
     r^2 \dot{\epsilon} = \sqrt{\Theta(\epsilon)}
\end{equation}
where the expressions for $R(r)$ and $\Theta(\epsilon)$ takes the form
\begin{equation}
    R(r) =  (n^2 - 1)E^2 (r^2 + a^2)^2 + \Big[(r^2 + a^2)E - aL_{\phi}\Big]^2 - \Delta\Big[(n^2 -1) a^2 E^2 + (aE - L_{\phi})^2 +\kappa\Big] 
\end{equation}
\begin{equation}
    \Theta(\epsilon) = \kappa~.  
\end{equation}
Since, shadows are formed by light rays moving in unstable circular orbits, so we use the conditions $R(r) = \frac{\partial R(r)}{\partial r}=0$. The conditions read
\begin{equation}
    (n^2 - 1)(r^2 + a^2)^2 + \Big[(r^2 + a^2) - a\xi\Big]^2 = \Delta\Big[(n^2 -1) a^2  + (a- \xi)^2 +\eta\Big]
\end{equation}
\begin{equation}
  4r(n^2 - 1)(r^2 + a^2) + 4r\Big[(r^2 + a^2) - a\xi\Big] -2nn^{'}\Delta a^2 +2(r^2 + a^2)nn^{'} = \Delta^{'}\Big[(n^2 -1) a^2  + (a- \xi)^2 +\eta\Big]~.  
\end{equation}
Eliminating $\eta$ from the above equations, we get a quadratic equation $\xi$ as
\begin{equation}
    A\xi^2 + 2B\xi + C = 0
\end{equation}
with the expressions for $A$, $B$ and $C$ taking the form
\begin{align}
    A = a^2 \Delta^{'}~~;~~ B = 2ar\Delta - a\Delta^{'}(r^2 + a^2)~;~~\\ \nonumber
    C = \Big[n^2 \Delta^{'}(r^2 + a^2)^2 -4r \Delta n^2 (r^2 + a^2) + 2nn^{'}\Delta^2 a^2 - 2(r^2 + a^2)^2 nn^{'}\Delta\Big]~.
\end{align}
Solving for $\xi$, we have
\begin{equation}
    \xi = -\frac{B}{A} \pm \sqrt{\Big(\frac{B}{A}\Big)^2 - \frac{C}{A}}
\end{equation}
with the negative sign yielding the appropriate results. The expression for $\eta$ becomes
\begin{equation}
    \eta = \frac{1}{\Delta}\Bigg[ (n^2 - 1)(r^2 + a^2)^2 + \Big[(r^2 + a^2) - a\xi\Big]^2 \Bigg] -(n^2 -1) a^2  - (a- \xi)^2~.
\end{equation}
The constants $\xi$ and $\eta$ are the quantities in terms of which shadow radius $R_s$ is evaluated. In our case, we consider the observer to be placed in the equatorial plane, that is, $\theta_0 = \frac{\pi}{2}$.

\noindent In order to determine $\alpha$ and $\beta$, we need the geodesic equations and thereby calculate $\Big(\frac{d \phi}{d r}\Big)$ and $\Big(\frac{d \epsilon}{d r}\Big)$. The geodesics for $\phi$, $\epsilon$ and $r$ have the form (considering near equatorial plane)
\begin{equation}
 \Big(\frac{d \phi}{d \lambda}\Big) = \frac{a}{\Delta r^2}\Big[(r^2 + a^2)E - aL_{\phi}\Big] + \frac{1}{r^2}\Big[L_{\phi} - a E \Big]
\end{equation}
\begin{equation}
\Big(\frac{d \epsilon}{d \lambda}\Big) = \frac{\sqrt{\kappa}}{r^2}   
\end{equation}
\begin{equation}
\Big(\frac{d r}{d \lambda}\Big) = \frac{1}{r^2}\sqrt{(n^2 - 1)E^2 (r^2 + a^2)^2 + \Big[(r^2 + a^2)E - aL_{\phi}\Big]^2 - \Delta\Big[(n^2 -1) a^2 E^2 + (aE - L_{\phi})^2 +\kappa\Big]}~.   
\end{equation}
Utilising the above geodesics, we get the relevant equations as
\begin{equation}
   \Big(\frac{d \phi}{d r}\Big) = \frac{\frac{a}{\Delta}\Big[(r^2 + a^2)E - aL_{\phi}\Big] + \Big[L_{\phi}\csc^2 \theta - a E \Big]}{\sqrt{(n^2 - 1)E^2 (r^2 + a^2)^2 + \Big[(r^2 + a^2)E - aL_{\phi}\Big]^2 - \Delta\Big[(n^2 -1) a^2 E^2 + (aE - L_{\phi})^2 +\kappa\Big]}}
\end{equation}
\begin{equation}
   \Big(\frac{d \epsilon}{d r}\Big) = \sqrt{\frac{\kappa}{(n^2 - 1)E^2 (r^2 + a^2)^2 + \Big[(r^2 + a^2)E - aL_{\phi}\Big]^2 - \Delta\Big[(n^2 -1) a^2 E^2 + (aE - L_{\phi})^2 +\kappa\Big]}}~.
\end{equation}
Now replacing the above relations in the expressions for $\alpha$ and $\beta$ and then taking the limit $r \to \infty$, we get the celestial coordinates as
\begin{equation}
    \alpha = -\frac{\xi}{n}~~;~~ \beta =\pm \frac{\sqrt{\eta}}{n}~.
\end{equation}
Plotting $\alpha$ along X-axis vs $\beta$ along Y-axis, we get the silhouette of the black hole shadow.\\

\noindent \textbf{Case IIb}\label{Ia}\\
Here, we consider $n(r)=$constant=$\sqrt{1-k}$. The $t$ and $\phi$ geodesics are the same as the previous case. In this case we find that the above eq.\eqref{19} is separable. Using the Cartar constant $\kappa$ \cite{66}, we have
\begin{align}
  \Big(\frac{\partial S_\theta}{\partial \theta}\Big)^2 + (n^2-1) a^2 E^2 \sin^2 \theta -a^2E^2\cos^2 \theta + L^2 _{\phi}\cot ^2 \theta = -\Delta \Big(\frac{\partial S_r}{\partial r}\Big)^2 + \frac{(n^2 - 1)}{\Delta}E^2 (r^2 + a^2)^2  \\ \nonumber +\frac{1}{\Delta}\Big[(r^2 + a^2)E + aL_{\phi}\Big]^2  - (aE - L_{\phi})^2   = \kappa ~.
\end{align}
This leads to the $r$ and $\theta$ equation(s) using \( \frac{\partial \mathcal{L}}{\partial \dot{x}^ \mu} = \frac{\partial S}{\partial x ^ \mu} \) ,
with \(\mathcal{L} = \frac{1}{2}g_{\mu \nu}\dot{x}^ \mu \dot{x}^ \nu \) as
\begin{equation}\label{003}
     \zeta^2 \dot{r} = \sqrt{R(r)}
\end{equation}
\begin{equation}\label{004}
     \zeta^2 \dot{\theta} = \sqrt{\Theta(\theta)}~.
\end{equation}
The expressions for $R(r)$ and $\Theta(\theta)$ take the form
\begin{equation}
    R(r) =  (n^2 - 1)E^2 (r^2 + a^2)^2 + \Big[(r^2 + a^2)E - aL_{\phi}\Big]^2 - \Delta\Big[(aE - L_{\phi})^2 +\kappa\Big] 
\end{equation}
\begin{equation}
    \Theta(\theta) = \kappa - (n^2-1) a^2 E^2 \sin^2 \theta + a^2E^2\cos^2 \theta - L^2 _{\phi}\cot ^2 \theta~.  
\end{equation}
The black hole shadow is formed by the unstable circular orbits of photons moving around black hole. The condition satisfied by these rays are $R(r)=\frac{\partial R}{\partial r}=0$. Imposing these conditions we get
\begin{equation}
    (n^2 - 1)(r^2 + a^2)^2 + \Big[(r^2 + a^2) - a\xi\Big]^2 = \Delta\Big[\eta + (a- \xi)^2 \Big]
\end{equation}
\begin{equation}
   4rn^2 (r^2 + a^2) -4ra\xi = \Delta^{'}\Big[\eta + (a- \xi)^2 \Big] 
\end{equation}
where $\xi = \frac{L_{\phi}}{E}$ and $\eta = \frac{\kappa}{E^2}$ are the Chandrasekhar constants \cite{66}. Eliminating $\eta$ from the above equation(s), we get an equation in $\xi$ of the form 
\begin{equation}
    A\xi^2 + 2B\xi + C = 0
\end{equation}
where, $A$, $B$, $C$ has the form
\begin{align}
    A = a^2 \Delta^{'}~~;~~ B = 2ar\Delta - a\Delta^{'}(r^2 + a^2)~;~~\\ \nonumber
    C = \Big[n^2 \Delta^{'}(r^2 + a^2)^2 -4r \Delta n^2 (r^2 + a^2)\Big]~.
\end{align}
Solving the above equation, $\xi$ becomes
\begin{equation}
    \xi = -\frac{B}{A} \pm \sqrt{\Big(\frac{B}{A}\Big)^2 - \frac{C}{A}}~.
\end{equation}
The shadow can now be obtained by considering the negative sign in the above solution. Also $\eta$ becomes
\begin{equation}
    \eta = \frac{1}{\Delta}\Bigg[ (n^2 - 1)(r^2 + a^2)^2 + \Big[(r^2 + a^2) - a\xi\Big]^2 \Bigg] - (a- \xi)^2~.
\end{equation}
In order to calculate the celestial coordinates ($\alpha, \beta$) we need to make use of the geodesics equations(s) for $r$, $\theta$ and $\phi$ as given in eq.(s) \eqref{003}, \eqref{004} and \eqref{002} respectively. Using them, we get the expressions for $\frac{d\phi}{dr}$ and $\frac{d\theta}{dr}$ as 
\begin{equation}
   \Big(\frac{d \phi}{d r}\Big) = \frac{\frac{a}{\Delta}\Big[(r^2 + a^2)E - aL_{\phi}\Big] + \Big[L_{\phi}\csc^2 \theta - a E \Big]}{\sqrt{(n^2 - 1)E^2 (r^2 + a^2)^2 + \Big[(r^2 + a^2)E - aL_{\phi}\Big]^2 - \Delta\Big[ (aE - L_{\phi})^2 +\kappa\Big]}}
\end{equation}
\begin{equation}
   \Big(\frac{d \theta}{d r}\Big) = \sqrt{\frac{\kappa - (n^2-1) a^2 E^2 \sin^2 \theta + a^2E^2\cos^2 \theta - L^2 _{\phi}\cot ^2 \theta}{(n^2 - 1)E^2 (r^2 + a^2)^2 + \Big[(r^2 + a^2)E - aL_{\phi}\Big]^2 - \Delta\Big[(aE - L_{\phi})^2 +\kappa\Big]}}~.
\end{equation}
Replacing the above eq.(s) in the expressions for $\alpha$ and $\beta$, we have
\begin{equation}
    \alpha = -\frac{\xi}{n}\csc \theta_0~~;~~ \beta =\pm \frac{\sqrt{\eta - (n^2-1) a^2 \sin^2 \theta_0 + a^2\cos^2 \theta_0 - \xi^2 \cot ^2 \theta_0}}{n}~.
\end{equation}
The shadow can be obtained by plotting $\alpha$ along $X$ axis and $\beta$ along $Y$ axis. The observer is fixed at infinity ($r_0=\infty$) and the position of the observer with respect to the direction of black hole spin ($a$) can be varied. So we consider three positions as $\theta_0 = \frac{\pi}{4}$, $\theta_0 = \frac{\pi}{3}$ and $\theta_0 = \frac{\pi}{2}$. The corresponding plots with variation in plasma parameter $k$ is shown in the next section \ref{sec5}.\\

\noindent Before ending this section, we want to mention that the general case (Case \ref{I}) in which $n$ is a function of both $r$ and $\theta$ boils down to the case case $n=\sqrt{1-k}$. This can be seen as follows. Setting $f_r (r)=\omega_c ^2 r^2$ and $f_{\theta} (\theta)=\omega_c ^2 a^2 \cos^2 \theta$, we get $\omega_p (r, \theta)= \omega_c $ = constant. Now we write the Hamiltonian ($\mathcal{H}$) in the form
\begin{equation}
    \mathcal{H}(x^{\mu}, p_{\mu})=\frac{1}{2}\Big[g^{\mu \nu}p_{\mu}p_{\nu} + \widetilde{\omega}_p(r,\theta) ^2\Big]
\end{equation}
where $\widetilde{\omega_p}$ is given by
\begin{equation}
   \widetilde{\omega}_p = \frac{\omega_p}{\sqrt{-g_{00}}}=\frac{\omega_c}{\sqrt{-g_{00}}}~.
\end{equation}
Now the refractive index is defined as \cite{65}
\begin{equation}
    n^2 (r, \theta)=1-\frac{ \widetilde{\omega}_p ^2}{\omega^2}= 1- \Big(\frac{\omega_c}{\omega_0}\Big)^2=1-k=  constant
\end{equation}
where we have defined $\Big(\frac{\omega_c}{\omega_0}\Big)^2 = k$ and used
$\omega=\omega_{0}/\sqrt{-g_{00}}$. Substituting $\omega_c$ in terms of $n$, we get 
\begin{eqnarray}
    \mathcal{H}(x^{\mu}, p_{\mu})&=&\frac{1}{2}\Big[g^{\mu \nu}p_{\mu}p_{\nu} -g^{00} \omega_c ^2\Big]\nonumber\\
    &=&\frac{1}{2}\Big[g^{\mu \nu}p_{\mu}p_{\nu} + (n^2 -1)g^{00} \omega_0 ^2\Big]~,~n=\sqrt{1-k}~.
\end{eqnarray}
This is the same Hamiltonian as in Case \ref{II} with $p_0 = -E=-\omega_0$.

\section{Impact of the spacetime, $PFDM$ and plasma parameters on the black hole shadow}\label{sec5}
The motion of any particle in the black hole spacetime is influenced by the parameters describing the spacetime. The same is true for unstable photons ($m=0$) that either plunge into the black hole singularity or fly off to infinity. \textbf{The photons} that fly off to infinity reach the observer and form the boundary of the black hole shadow. Thus the shadow formed by photons gets impacted by the spacetime parameters.

\noindent The parameters describing the spacetime are spin ($a$) and charge ($Q$) of the black hole, $PFDM$ parameter ($\chi$) and the plasma \textbf{parameters $k$ and $\frac{\omega_c}{\omega_0}$ }. Now we show the plots and discuss how the parameters affect the black hole shadow. 
\begin{figure}[H]
  \centering
  \begin{minipage}[b]{0.35\textwidth}
   \subfloat[\footnotesize Q = 0.2, k = 0.2, $\chi$ = 0.2 ]{\includegraphics[width=\textwidth]{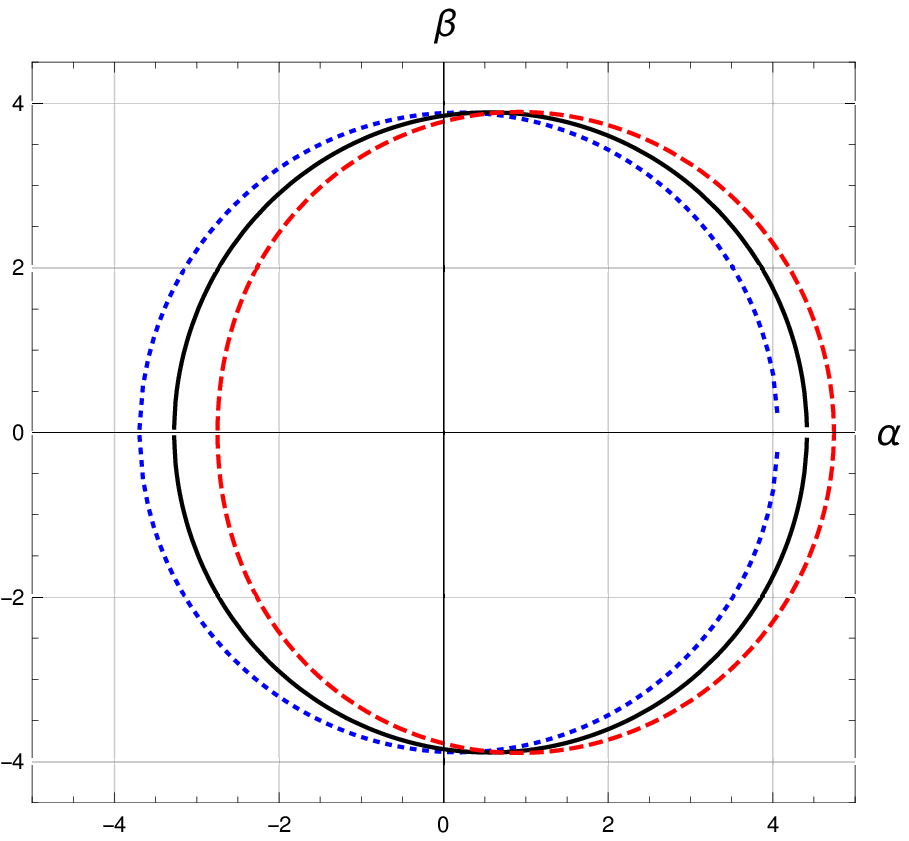}}
    \end{minipage}
  \hspace{1.0cm}
   \begin{minipage}[b]{0.35\textwidth}
    \subfloat[\footnotesize  Q = 0.2, k = 0.2, $\chi$ = 1.0 ]{\includegraphics[width=\textwidth]{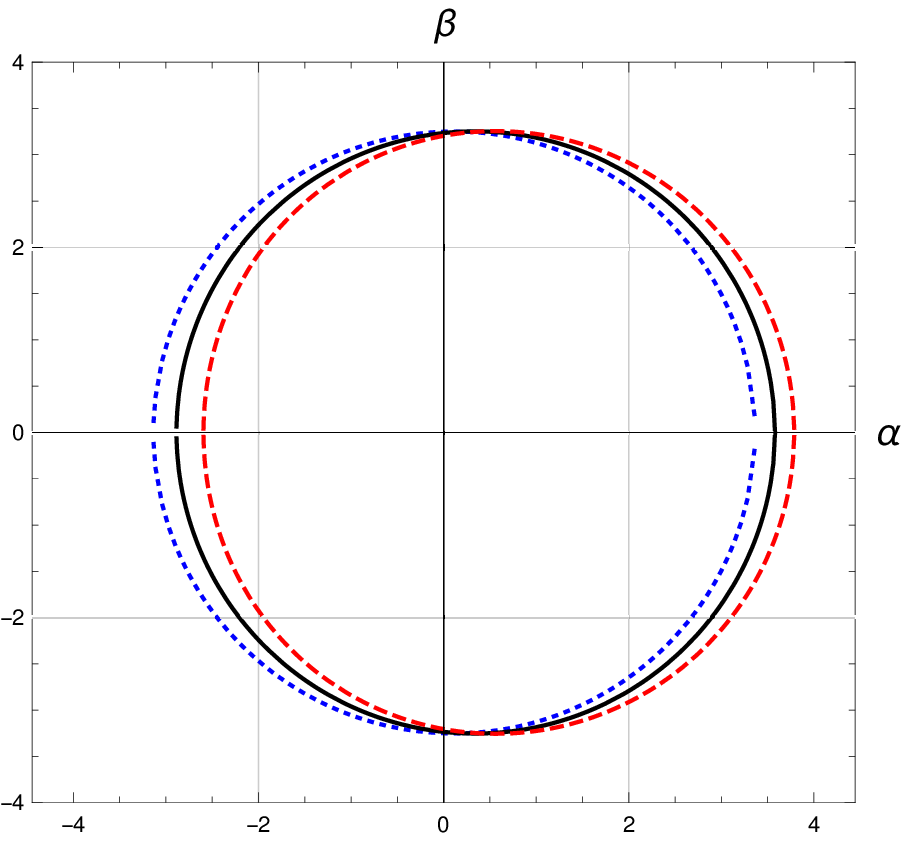}}
     \end{minipage}
  \caption{\footnotesize Variation of the black hole shadow with spin ($a$) of the black hole. The colored plots are for different spin values-blue dotted ($a=0.1$), black ($a=0.3$), red dashed ($a=0.5$). The plots are shown for inhomogeneous plasma with $n(r)=\sqrt{1-\frac{k}{r}}.$ }
  \label{2}
\end{figure}

\noindent In the Fig.\ref{2}, we have shown the impact of the black hole spin ($a$) on the shadow. In order to do so, we have fixed the rest of the black hole parameters ($Q$ = 0.2, $\chi$ = 0.2, 1.0, $k$ = 0.2). The plots are shown for inhomogeneous plasma with $n(r)=\sqrt{1-\frac{k}{r}}.$ Previously, we have mentioned that $PFDM$ parameter $\chi$ has two ranges of values separated by $\chi_c$. The left plot is for $\chi < \chi_c$ and the right one is for $\chi > \chi_c$. The value of $\chi_c$ varies with variation in the values of black hole spin $a$ with fixed charge $Q$ as can be seen in Table \ref{Fig11}. The shadow is larger in case of $\chi < \chi_c$, whereas, they are comparatively smaller in case of $\chi > \chi_c$. Besides, we find that with the increase in spin ($a$) of the black hole, the shadow gets rotated and slightly deformed. This is due to the rotational drag force on the unstable photons moving in close vicinity of the black hole.

\begin{figure}[H]
  \centering
  \begin{minipage}[b]{0.3\textwidth}
   \subfloat[\footnotesize a = 0.4, k = 0.2, $\chi$ = 0.2 ]{\includegraphics[width=\textwidth]{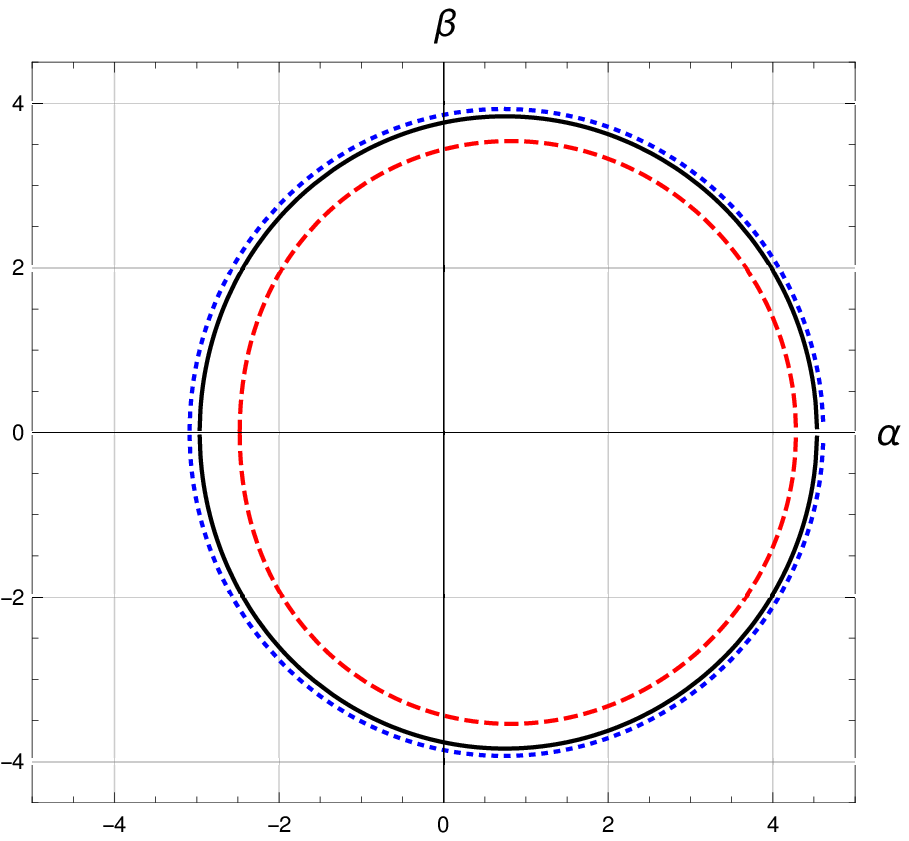}}
    \end{minipage}
  \hspace{1.0cm}
   \begin{minipage}[b]{0.3\textwidth}
    \subfloat[\footnotesize  a = 0.4, k = 0.2, $\chi$ = 1.0 ]{\includegraphics[width=\textwidth]{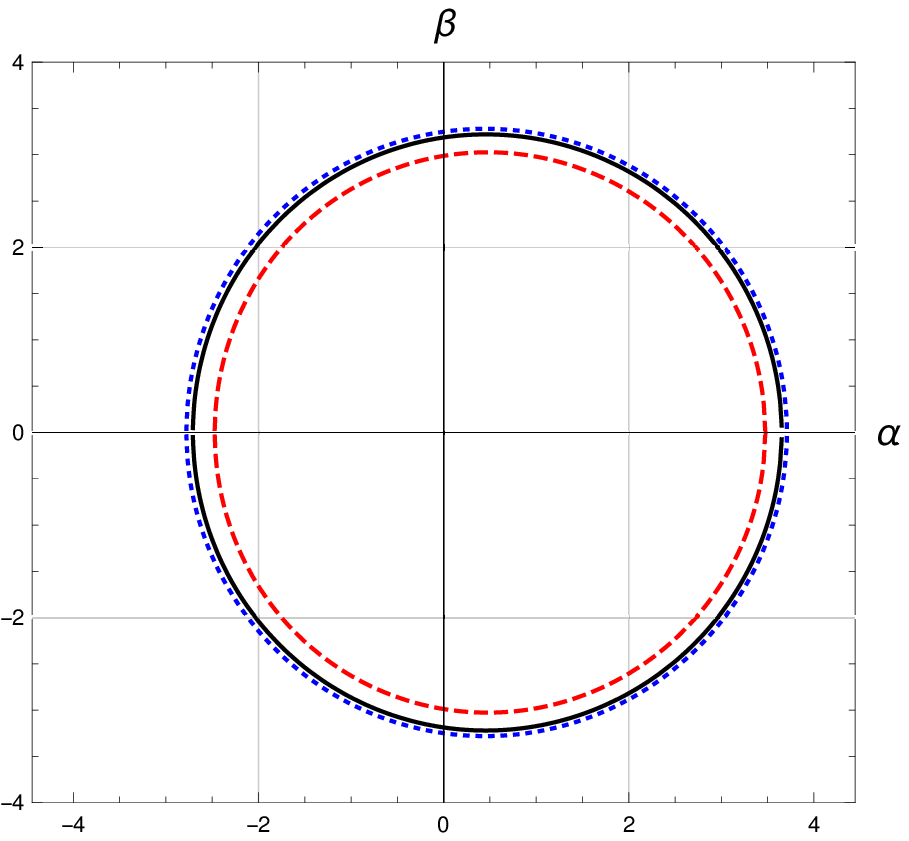}}
     \end{minipage}
  \caption{\footnotesize Variation of the black hole shadow with charge $Q$. The colored plots are for different charge values-blue dotted ($Q=0.0$), black ($Q=0.3$), red dashed ($Q=0.6$). The plots are shown for inhomogeneous plasma with $n(r)=\sqrt{1-\frac{k}{r}}.$}
  \label{1}
\end{figure}

\noindent Fig.\ref{1} depict the effect of charge ($Q$) on the shadow of the black hole. Just like the previous case we have shown two ranges of $\chi$ ($\chi < \chi_c$ and $\chi > \chi_c$). We set the rest of the black hole parameters to constant values ($a$ = 0.4, $\chi$ = 0.2, 1.0, $k$ = 0.2). The plots are shown for inhomogeneous plasma with $n(r)=\sqrt{1-\frac{k}{r}}.$ In this case too, the value of $\chi_c$ varies with variation in the values of black hole charge $Q$ with fixed spin $a$ as can be seen in Table \ref{Fig11}. Here too, we find that the shadow size in $\chi < \chi_c$ is greater than that in $\chi > \chi_c$. Also with the increase in charge ($Q$) of the black hole, the shadow size reduces in both cases ($\chi < \chi_c$ and $\chi > \chi_c$). The reason for the observation can be assigned to the fact that the event horizon radius ($r_{h+}=M+\sqrt{M^2 - Q^2}$) without dark matter ($\chi$=0) and plasma ($k$=0) decreases with increase in charge $Q$. The same persists in presence of $\chi$ and $k$. The black hole shadow which is a manifestation of the event horizon, thereby decreases. The decrement in shadow size with charge ($Q$) is non-uniform. 
\begin{figure}[H]
  \centering
  \begin{minipage}[b]{0.35\textwidth}
   \subfloat[\footnotesize  a=0.4, Q = 0.2, k = 0.2,  $\chi < \chi_c$ ]{\includegraphics[width=\textwidth]{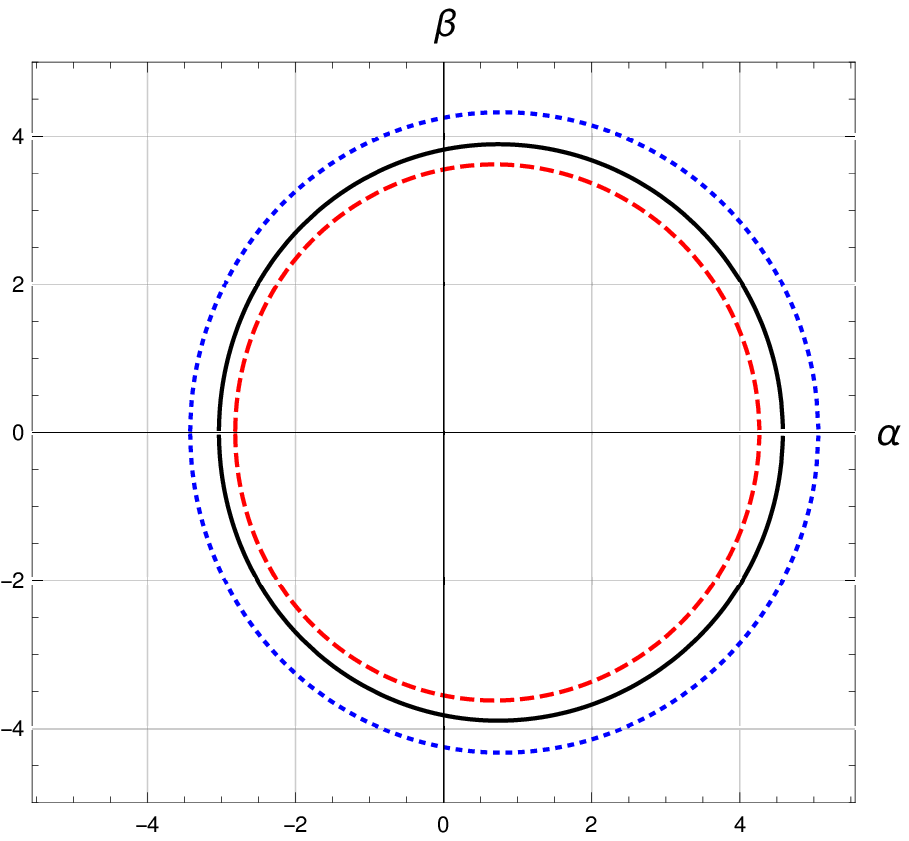}}
    \end{minipage}
  \hspace{1.0cm}
   \begin{minipage}[b]{0.35\textwidth}
    \subfloat[\footnotesize  a=0.4, Q = 0.2, k = 0.2, $\chi > \chi_c$  ]{\includegraphics[width=\textwidth]{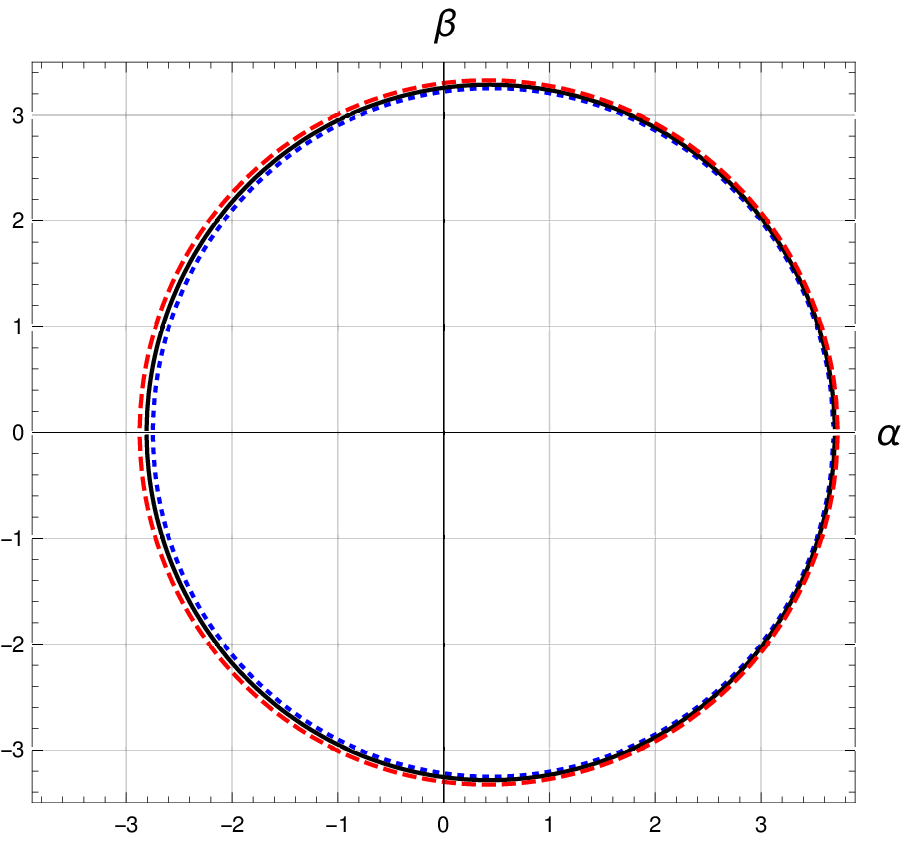}}
     \end{minipage}
  \caption{ \footnotesize Variation of black hole shadow with perfect fluid dark matter ($PFDM$) parameter $\chi$. The colored plots are for different values of $\chi$-blue dotted ($\chi=0.1$), black ($\chi=0.2$), red dashed ($\chi=0.3$) for the left plot and blue dotted ($\chi=1.0$), black ($\chi=1.1$), red dashed ($\chi=1.2$) for the right plot. The plots are shown for inhomogeneous plasma with $n(r)=\sqrt{1-\frac{k}{r}}.$}
  \label{3}
\end{figure}

\noindent In Fig.\ref{3}, we have shown the impact of $PFDM$ parameter ($\chi$) on the black hole shadow. The plots are shown for inhomogeneous plasma with $n(r)=\sqrt{1-\frac{k}{r}}.$ From previous analysis, we find that the outer event horizon radius ($r_{h+}$) decreases with increase in $\chi$ for $\chi < \chi_c$ and increases for  $\chi > \chi_c$. The analogical results are observed in case of black hole shadow. We found that for  $\chi < \chi_c$, the shadow decreases non-uniformly and gets distorted with increase in $\chi$. On the other hand, for  $\chi > \chi_c$, the shadow increases uniformly with 
\begin{figure}[H]
  \centering
  \begin{minipage}[b]{0.3\textwidth}
   \subfloat[\footnotesize  a=0.4, Q = 0.2, $\chi$ = 0.2 ]{\includegraphics[width=\textwidth]{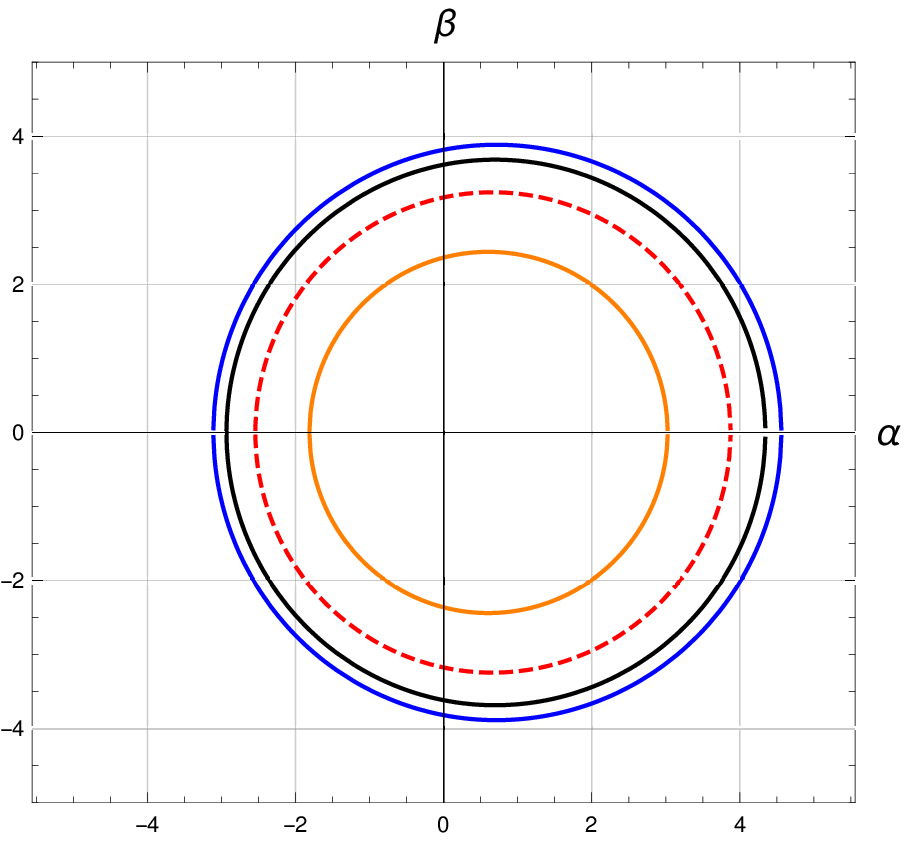}}
    \end{minipage}
    \hspace{1.0cm}
   \begin{minipage}[b]{0.3\textwidth}
    \subfloat[\footnotesize  a=0.4, Q = 0.2, $\chi$ = 1.0 ]{\includegraphics[width=\textwidth]{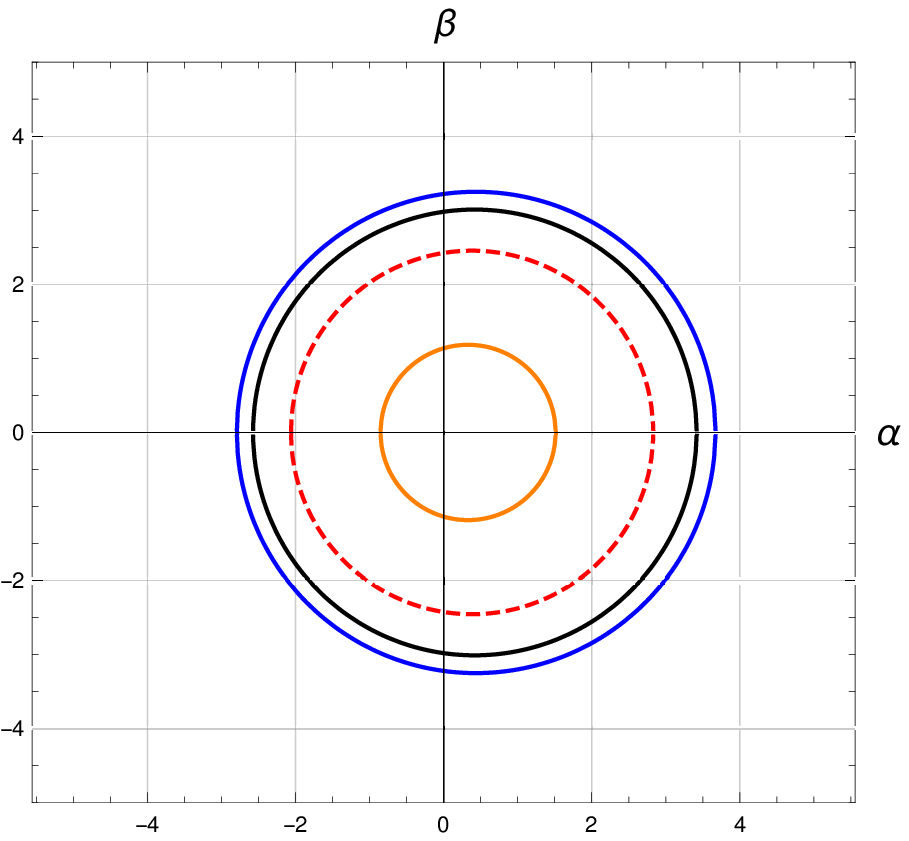}}
     \end{minipage}
  \caption{\footnotesize Variation of black hole shadow with $f_r (r)= \omega_c ^2 \sqrt{M^3 r} $ and $f_{\theta}(\theta) =0$ for $\chi=0.2$ (left) and $\chi=1.0$ (right). The plots are shown for $M=1$ and different values of $\Big(\frac{\omega_c}{\omega_0}\Big)^2$ with $\Big(\frac{\omega_c}{\omega_0}\Big)^2=0.0$ (blue), $\Big(\frac{\omega_c}{\omega_0}\Big)^2=1.0$ (black), $\Big(\frac{\omega_c}{\omega_0}\Big)^2=3.0$ (red dotted) and $\Big(\frac{\omega_c}{\omega_0}\Big)^2=6.0$ (orange).}
  \label{10a}
\end{figure}

\noindent  increase in $\chi$, though the effect is less pronounced than for  $\chi < \chi_c$. Such kind of observation results from the fact that $PFDM$ effectively gives the mass of the total system as discussed previously.

\noindent In Figure \ref{10a}, we have shown the variation of black hole shadow with a certain form of the function $f_r (r) = \omega_c ^2 \sqrt{r}$ and $f_{\theta}(\theta)=0$, with $M=1$. The plots are shown for black hole spin $a=0.4$ and charge $Q=0.2$. Also the plots are shown with the observer in the equatorial plane, $\theta_0 = \frac{\pi}{2}$. The left one is for dark matter parameter $\chi=0.2$ and the right one for $\chi=1.0.$ We have taken these two values since $\chi=0.2$ is less than $\chi_c$ whereas,  $\chi=1.0$ is greater than $\chi_c$ for the considered combination of spin $a$ and charge $Q$. The plots show that shadow size decreases with increase in plasma parameter $\omega_c$. Besides, we also observe that the shadow size is smaller in case of $\chi=1.0$ as compared to that in $\chi=0.2$. 

\begin{figure}[H]
  \centering
  \begin{minipage}[b]{0.35\textwidth}
   \subfloat[\footnotesize  a=0.4, Q = 0.2, $\chi$ = 0.2 ]{\includegraphics[width=\textwidth]{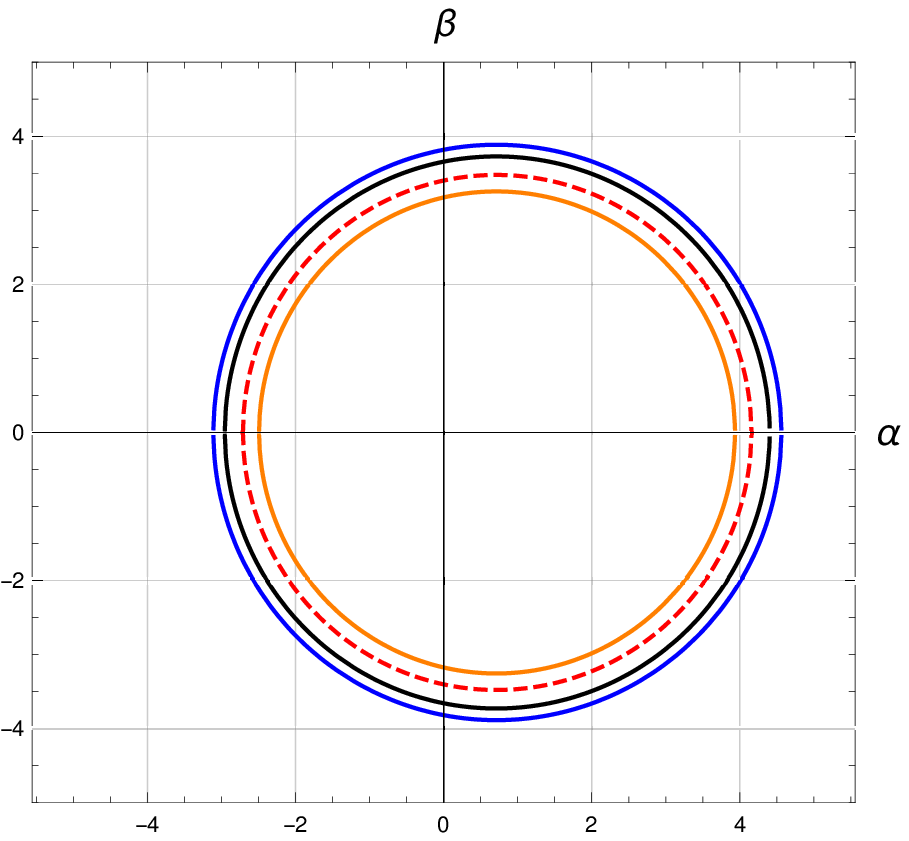}}
    \end{minipage}
    \hspace{1.0cm}
   \begin{minipage}[b]{0.35\textwidth}
    \subfloat[\footnotesize  a=0.4, Q = 0.2, $\chi$ = 1.0 ]{\includegraphics[width=\textwidth]{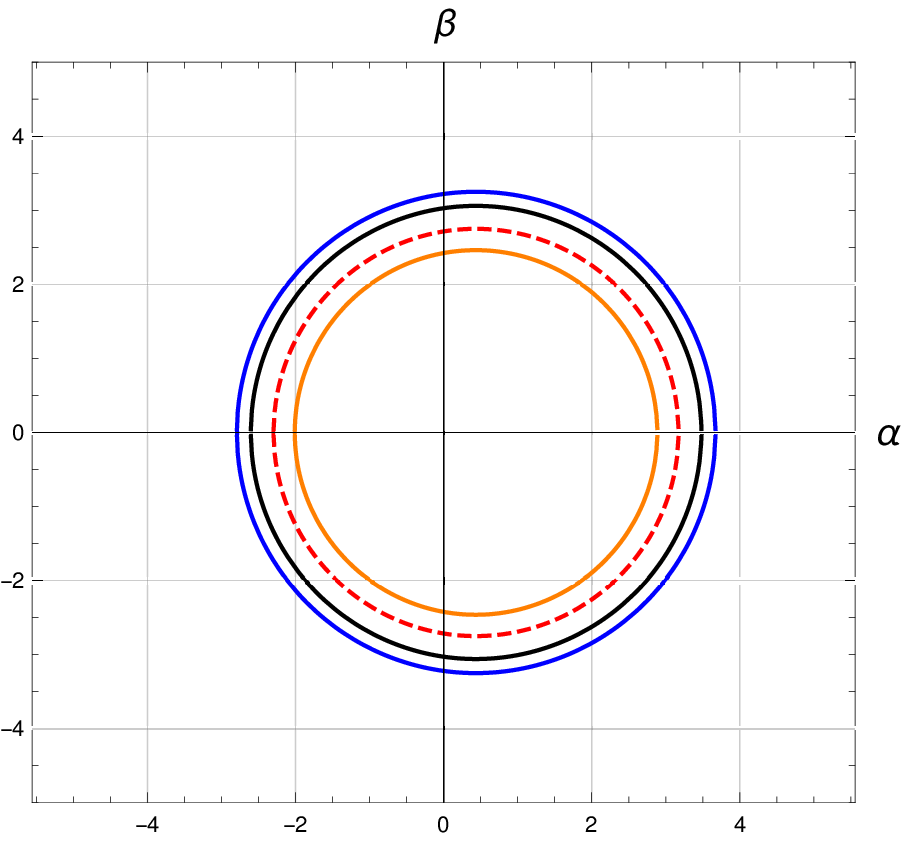}}
     \end{minipage}
  \caption{\footnotesize Variation of black hole shadow with $f_r (r)= 0 $ and $f_{\theta}(\theta) = \omega_c ^2 M^2 \Big(1 + 2 \sin^2 \theta\Big)$ for $\chi=0.2$ (left) and $\chi=1.0$ (right). The plots are shown for $M=1$ and different values of $\Big(\frac{\omega_c}{\omega_0}\Big)^2$ with $\Big(\frac{\omega_c}{\omega_0}\Big)^2=0.0$ (blue), $\Big(\frac{\omega_c}{\omega_0}\Big)^2=0.4$ (black),$\Big(\frac{\omega_c}{\omega_0}\Big)^2=1.0$ (red dotted) and $\Big(\frac{\omega_c}{\omega_0}\Big)^2=1.5$ (orange).}
  \label{10b}
\end{figure}

\noindent In Figure \ref{10b}, we have shown the variation of black hole shadow with $f_r (r) = 0$ and $f_{\theta}(\theta)= \omega_c ^2 \Big(1 + 2 \sin^2 \theta\Big)$, with $M=1$. The plots are shown with spin and charge of the black hole set at $a=0.4$ and $Q=0.2$ respectively. Also, the plots are shown with the observer situated in the equatorial plane, $\theta_0 = \frac{\pi}{2}$. The left plot is for dark matter parameter $\chi=0.2$ and the right one is for $\chi=1.0.$  The plots show that shadow size once again decreases with increase in plasma parameter $\omega_c$  as in the earlier case.

\begin{figure}[H]
  \centering
  \begin{minipage}[b]{0.4\textwidth}
   \subfloat[\footnotesize  a=0.4, Q = 0.2, $\chi$ = 0.2 ]{\includegraphics[width=\textwidth]{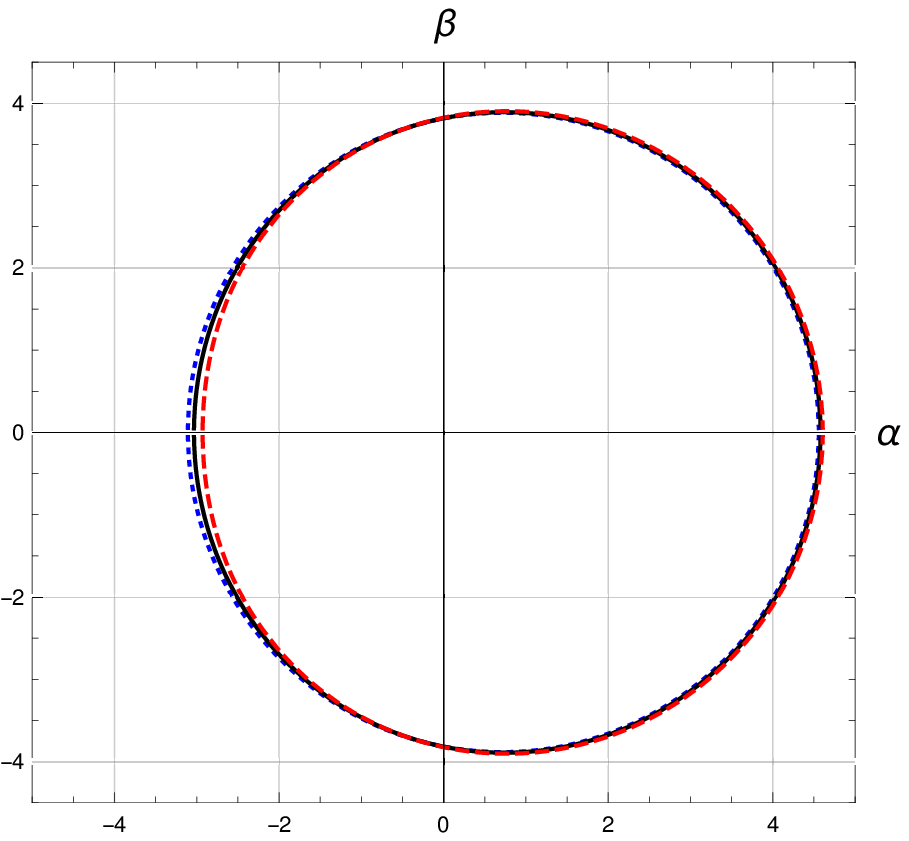}}
    \end{minipage}
    \hspace{1.0cm}
   \begin{minipage}[b]{0.4\textwidth}
    \subfloat[\footnotesize  a=0.4, Q = 0.2, $\chi$ = 1.0 ]{\includegraphics[width=\textwidth]{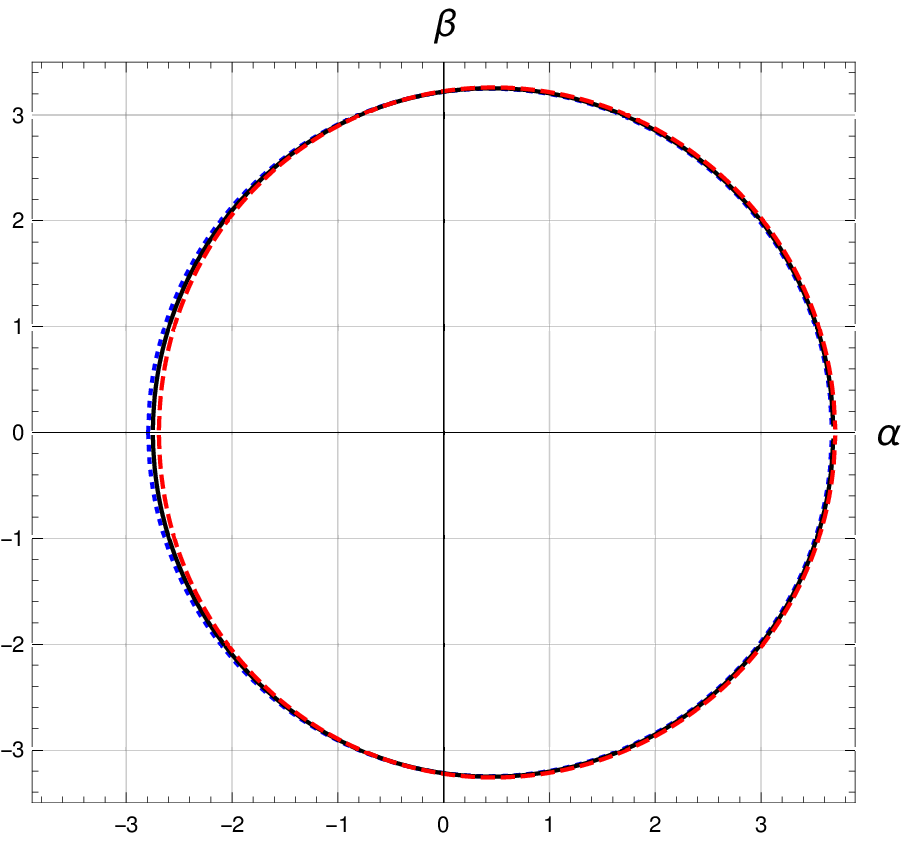}}
     \end{minipage}
  \caption{\footnotesize Variation of the black hole shadow with inhomogeneous plasma \Big($n = n(r)=\sqrt{1-(\frac{k}{r}})$\Big). The colored plots are for different values of plasma parameter-blue dotted ($k=0.0$), black ($k=0.2$), red dashed ($k=0.4$).}
  \label{5}
\end{figure}

\noindent We plot the effect of inhomogeneous plasma ($n = n(r)$) on the black hole shadow in Fig.\ref{5}. We show the plots both for $\chi < \chi_c$ and $\chi > \chi_c$. We observe that the co-rotating photon radius ($r_{p1}$) which correspond to the extreme left of $\alpha$ axis decreases with increase in plasma parameter $k$. The same happens in case of counter rotating radius ($r_{p2}$) which corresponds to the extreme right of the $\alpha$ axis. The effect is the same as obtained previously following a numerical approach. The cumulative effect of the two extreme orbit produces the unstable photon orbit which forms the black hole shadow. The effect remains identical for both $\chi < \chi_c$ and $\chi > \chi_c$. The shadow size is larger in case of $\chi < \chi_c$ than that in $\chi > \chi_c$.

\begin{figure}[H]
  \centering
  \begin{minipage}[b]{0.4\textwidth}
   \subfloat[\footnotesize  a=0.4, Q = 0.2, $\chi$ = 0.2]{\includegraphics[width=\textwidth]{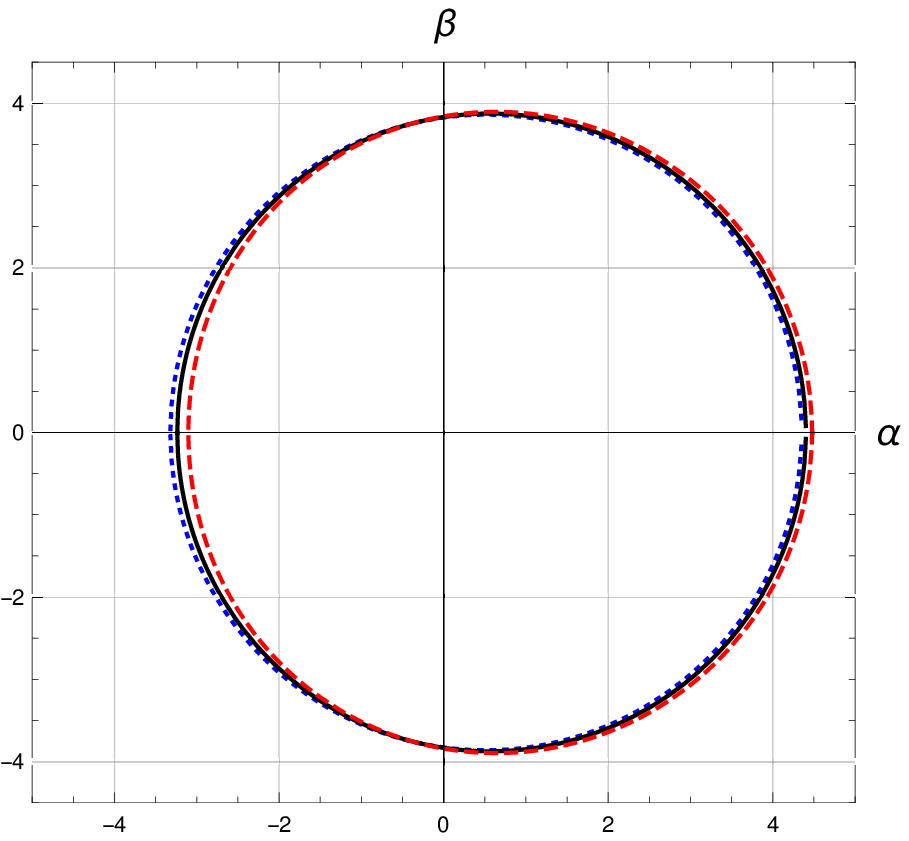}}
    \end{minipage}
  \hspace{1.0cm}
   \begin{minipage}[b]{0.4\textwidth}
    \subfloat[\footnotesize  a=0.4, Q = 0.2, $\chi$ =1.0 ]{\includegraphics[width=\textwidth]{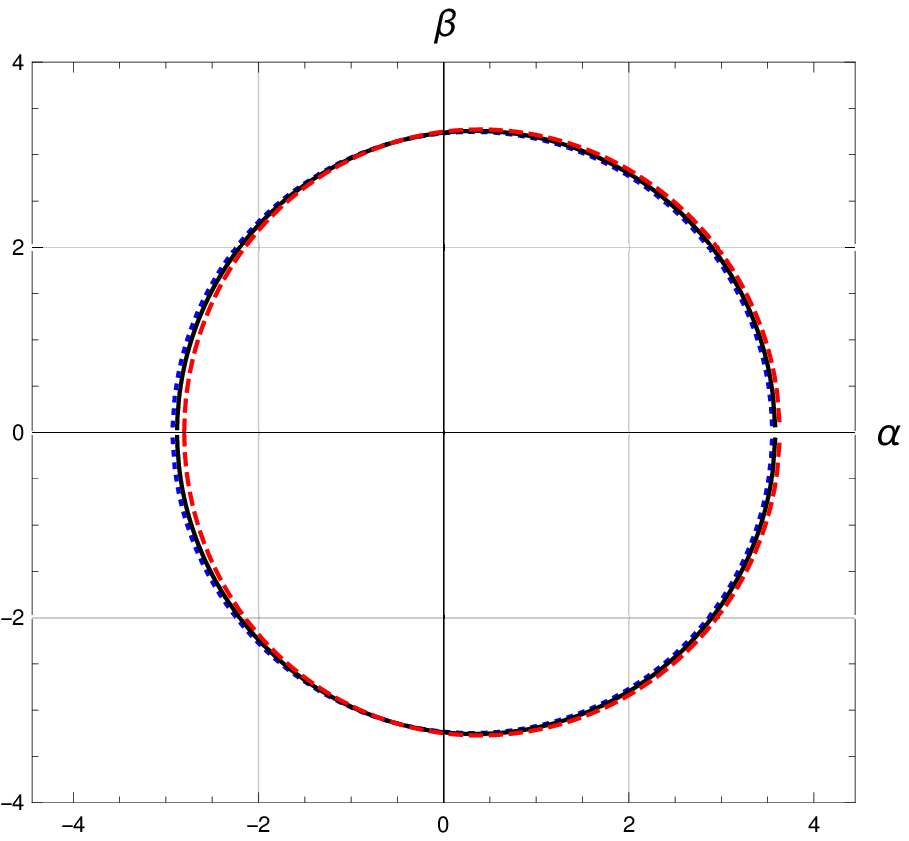}}
     \end{minipage}
  \caption{\footnotesize Variation of the black hole shadow with homogeneous plasma ($n =\sqrt{1-k}= $constant). The colored plots are for different values of plasma parameter-blue dotted ($k=0.0$), black ($k=0.2$), red dashed ($k=0.4$). The plots are for $\theta_0 =\frac{\pi}{4}$.}
  \label{4a}
\end{figure}
\begin{figure}[H]
  \centering
  \begin{minipage}[b]{0.4\textwidth}
   \subfloat[\footnotesize  a=0.4, Q = 0.2, $\chi$ = 0.2]{\includegraphics[width=\textwidth]{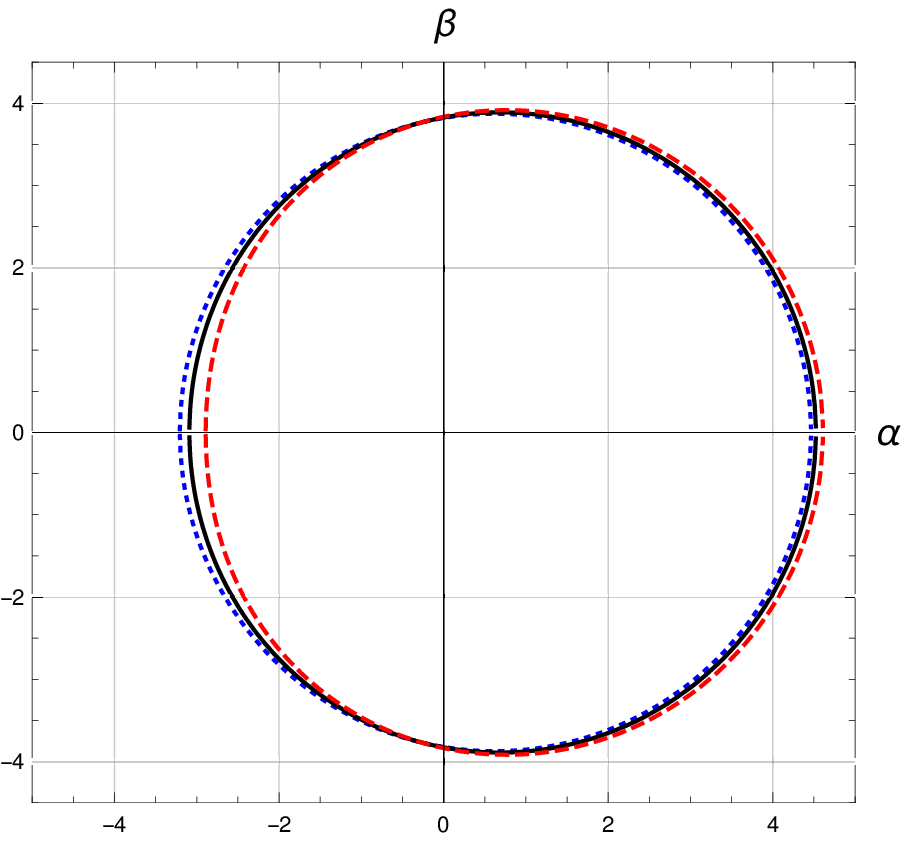}}
    \end{minipage}
  \hspace{1.0cm}
   \begin{minipage}[b]{0.4\textwidth}
    \subfloat[\footnotesize  a=0.4, Q = 0.2, $\chi$ =1.0 ]{\includegraphics[width=\textwidth]{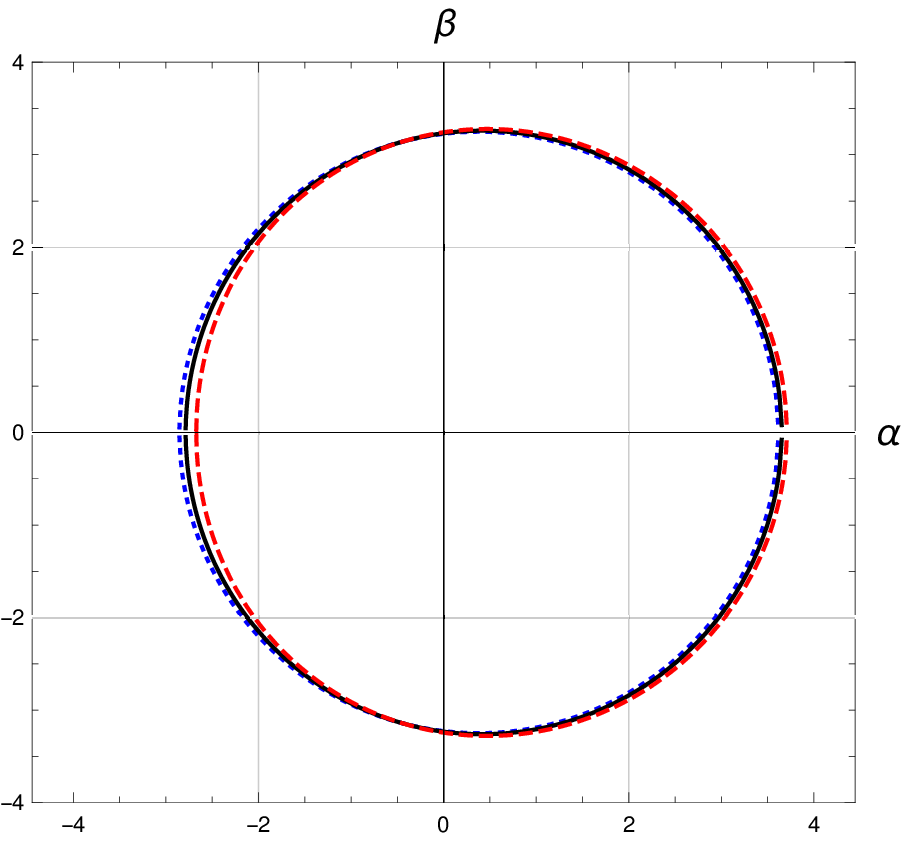}}
     \end{minipage}
  \caption{\footnotesize Variation of the black hole shadow with homogeneous plasma ($n =\sqrt{1-k}= $constant). The colored plots are for different values of plasma parameter-blue dotted ($k=0.0$), black ($k=0.2$), red dashed ($k=0.4$). The plots are for $\theta_0 =\frac{\pi}{3}$.}
  \label{4b}
\end{figure}
\begin{figure}[H]
  \centering
  \begin{minipage}[b]{0.4\textwidth}
   \subfloat[\footnotesize  a=0.4, Q = 0.2, $\chi$ = 0.2]{\includegraphics[width=\textwidth]{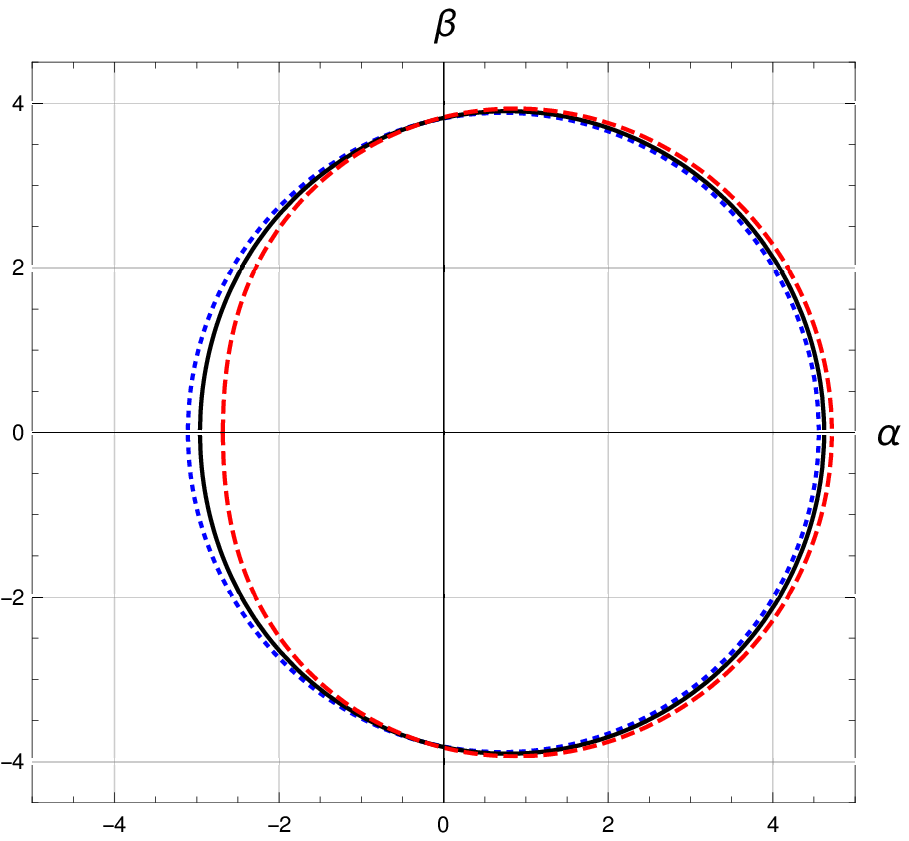}}
    \end{minipage}
  \hspace{1.0cm}
   \begin{minipage}[b]{0.4\textwidth}
    \subfloat[\footnotesize  a=0.4, Q = 0.2, $\chi$ =1.0 ]{\includegraphics[width=\textwidth]{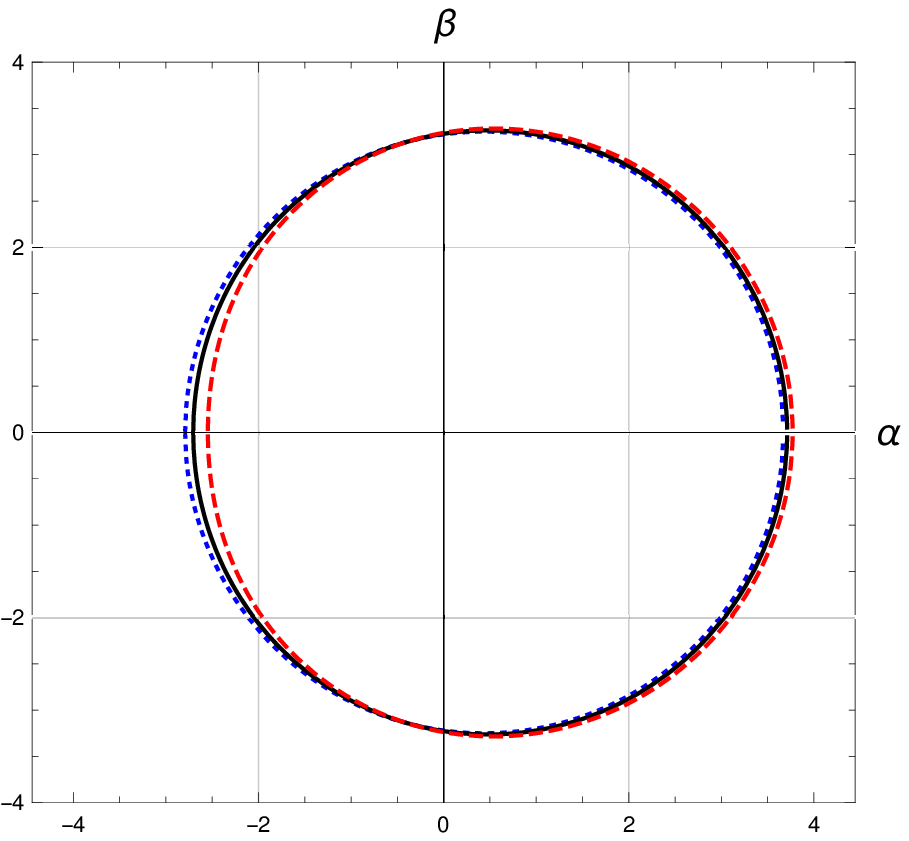}}
     \end{minipage}
  \caption{\footnotesize Variation of the black hole shadow with homogeneous plasma ($n =\sqrt{1-k}= $constant). The colored plots are for different values of plasma parameter-blue dotted ($k=0.0$), black ($k=0.2$), red dashed ($k=0.4$). The plots are for $\theta_0 =\frac{\pi}{2}$.}
  \label{4c}
\end{figure}
\noindent The effect of plasma is being observed for both homogeneous and inhomogeneous plasma distribution. In Fig.(s)(\ref{4a}, \ref{4b}, \ref{4c}), we have shown the effect of homogeneous plasma  ($n$=constant) on the black hole shadow. The observation is carried out for $\theta_0 = \frac{\pi}{4}$, $\theta_0 = \frac{\pi}{3}$ and $\theta_0 = \frac{\pi}{2}$ respectively. The extreme right point on the $\alpha$ axis corresponds to the radius of counter rotating photon orbits \cite{5a}. On the other hand, the extreme left point on the $\alpha$ axis corresponds to the radius of co-rotating orbit. The radius ($r_{p1}$) of co-rotating photons is found to decrease with increase in plasma parameter $k$, whereas that of counter rotating photons ($r_{p2}$) is observed to increase with increase in $k$.  The same is observed both for $PFDM$ parameter $\chi = 0.2$ and $1.0$. Also the shadow size is larger in case of $\chi =0.2$ with respect to that in $\chi = 1.0$.

\section{Effective potential ($V_{eff}$)}\label{sec6}
In this section, we study the effective potential ($V_{eff}$) as faced by a photon moving in the black hole spacetime. The potential can have maxima or minima which corresponds to the existence of unstable or stable orbits. The condition for maxima or minima are given as $\frac{\partial ^2 V_{eff}}{\partial r^2}<0$  and $\frac{\partial ^2 V_{eff}}{\partial r^2}>0$  respectively. The effective potential can be obtained from the modified radial equation which gives
\begin{equation}\label{53}
   \dot{r}^2 + V_{eff}=E^2 
\end{equation}
with the effective potential given by 
\begin{eqnarray}
    V_{eff}=-\frac{(a^2 E^2 -L_{\phi}^2)}{r^2} - \frac{2M}{r^3}(aE - L_{\phi})^2 + \frac{Q^2}{r^4}(aE-L_{\phi})^2 + \frac{\chi}{r^3}(aE - L_{\phi})^2  \\
    -(n^2 - 1)\Bigg[E^2 + \frac{a^2 E^2}{r^2} + a^2 E^2 \Big(\frac{2M}{r^3} - \frac{Q^2}{r^4} - \frac{\chi}{r^3}\ln{\frac{r}{|\chi|}}\Big)\Bigg]~.
\end{eqnarray}
The plots for the effective potential are shown below. Here we basically focus on the dependence of the effective potential on the plasma parameter $k$.
\begin{figure}[H]
  \centering
  \begin{minipage}[b]{0.45\textwidth}
   \subfloat[\footnotesize  $a$=0.5, $Q$ = 0.3, $\chi$ = 0.2, $L_{\phi}$=3.0 ]{\includegraphics[width=\textwidth]{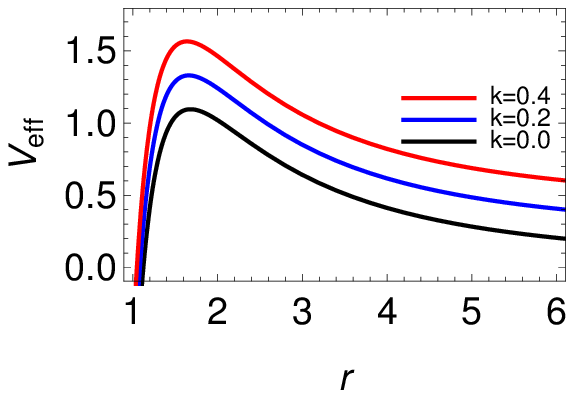}}
    \end{minipage}
    \hspace{1.0cm}
   \begin{minipage}[b]{0.45\textwidth}
    \subfloat[\footnotesize  $a$=0.5, $Q$ = 0.3, $\chi$ = 1.0, $L_{\phi}$=3.0 ]{\includegraphics[width=\textwidth]{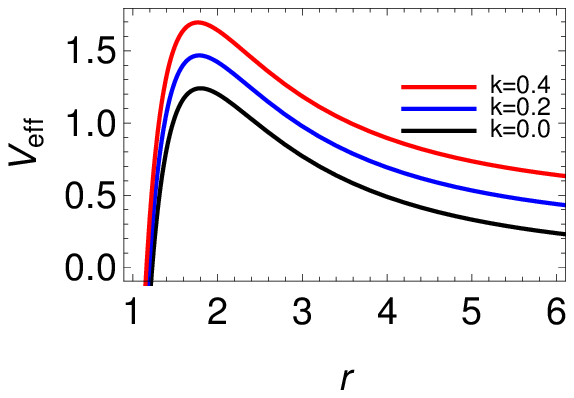}}
     \end{minipage}
  \caption{\footnotesize Variation of the effective potential ($V_{eff}$) for co-rotating photons with homogeneous plasma ($n = $constant$=\sqrt{1-k}$).}
  \label{6}
\end{figure}

\noindent The above Fig.\ref{6} shows the effective potential ($V_{eff}$) encountered by the photons in co-rotating orbits with variation in plasma parameter $k$. Here we consider the plasma distribution to be homogenous such that $n=\sqrt{1-k}$. The left one is for $\chi = 0.2$ and the right one for $\chi = 1.0$. The plots are shown by considering $M=1$, $E=1$, $a=0.5$, $Q=0.3$, $L_{\phi}$=3.0. We find that with increase in plasma parameter the potential increases uniformly in both cases. The potential shows a maxima which corresponds to unstable photon orbits. The maxima in case of $\chi=1.0$ are a little higher than the same for $\chi=0.2$. Also we find that the position of the maxima, which gives the unstable photon radius ($r_p$)  slightly shifts towards left with increase in plasma parameter $k$.
\begin{figure}[H]
  \centering
  \begin{minipage}[b]{0.45\textwidth}
   \subfloat[\footnotesize  $a$=0.5, $Q$ = 0.3, $\chi$ = 0.2, $L_{\phi}$=3.0 ]{\includegraphics[width=\textwidth]{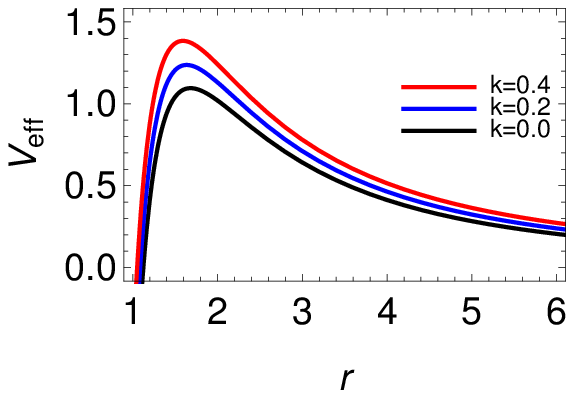}}
    \end{minipage}
    \hspace{1.0cm}
   \begin{minipage}[b]{0.45\textwidth}
    \subfloat[\footnotesize  $a$=0.5, $Q$ = 0.3, $\chi$ = 1.0, $L_{\phi}$=3.0 ]{\includegraphics[width=\textwidth]{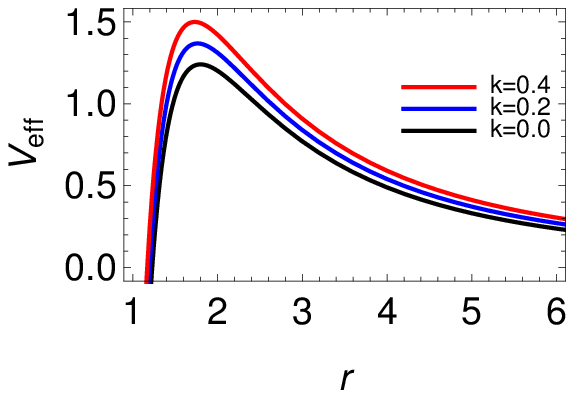}}
     \end{minipage}
  \caption{\footnotesize Variation of the effective potential ($V_{eff}$) for co-rotating photons with inhomogeneous plasma ($n = n(r)=\sqrt{1-\frac{k}{r}}$).}
  \label{7}
\end{figure}
\noindent Fig.\ref{7} shows the effective potential ($V_{eff}$) faced by the co-rotating photons with variation in plasma parameter $k$. The left one is for $\chi = 0.2$ and the right one for $\chi = 1.0$. We consider the plasma distribution to be inhomogeneous such that $n(r)=\sqrt{1-\frac{k}{r}}$. The plots are shown by considering $M=1$, $E=1$, $a=0.5$, $Q=0.3$, $L_{\phi}$=3.0. We find that with increase in plasma parameter ($k$), the potential increases uniformly in both cases. The potential shows a maxima which corresponds to unstable photon orbits. The maxima in case of $\chi=1.0$ are a little higher than the same for $\chi=0.2$. Also we find that the position of the maxima shifts towards left with increase in plasma parameter $k$. This implies that the radius ($r_p$) of the unstable photon orbits decreases with increase in plasma parameter $k$, i.e., the orbits move close to the black hole. The increment of effective potential in both the above cases for homogeneous and inhomogeneous plasma can be assigned to the fact that due to interaction of photons with plasma, the total energy and thereby the potential of the system increases. This can be seen by looking at the Hamiltonian ($\mathcal{H}$) which has an extra term due to plasma (eq.\eqref{223})
\begin{equation}
\begin{split}
    \mathcal{H_{I}} & =-\frac{1}{2} (n^2 -1)\Big(p_0 \sqrt{-g^{00}}\Big)^2 \\
    & =\frac{1}{2}\frac{k}{r^h}\Big(p_0 \sqrt{-g^{00}}\Big)^2 ~~;~~ n=\sqrt{1-\frac{k}{r^h}}~.
\end{split}
\end{equation}
\begin{figure}[H]
  \centering
  \begin{minipage}[b]{0.45\textwidth}
   \subfloat[\footnotesize  $a$=0.5, $Q$ = 0.3, $\chi$ = 0.2, $k$=0.2 ]{\includegraphics[width=\textwidth]{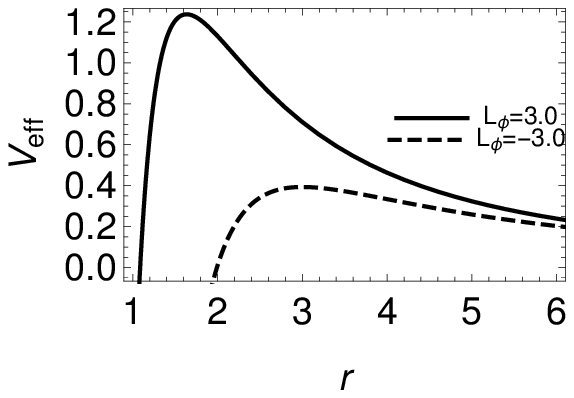}}
    \end{minipage}
    \hspace{1.0cm}
   \begin{minipage}[b]{0.45\textwidth}
    \subfloat[\footnotesize  $a$=0.5, $Q$ = 0.3, $\chi$ = 1.0, $k$=0.2 ]{\includegraphics[width=\textwidth]{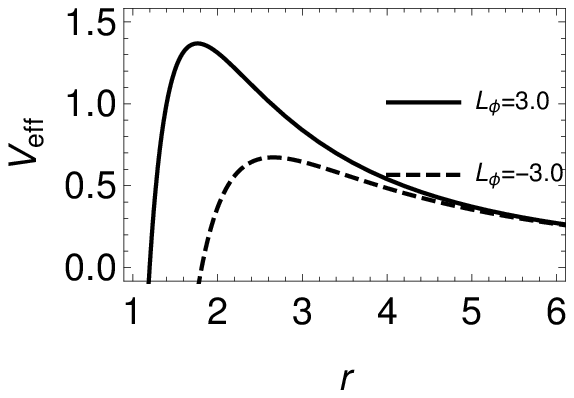}}
     \end{minipage}
  \caption{\footnotesize Variation of the effective potential ($V_{eff}$) in inhomogeneous plasma \Big($n = n(r)=\sqrt{1-\frac{k}{r}}$\Big) for $L_{\phi} >0$ and $L_{\phi}<0$. The solid line corresponds to corotating and the dashed line corresponds to the counter rotating orbit.}
  \label{8}
\end{figure}

\noindent Thus, with increase in plasma parameter $k$, the interaction energy increases. So by radial equation \eqref{53} for fixed $r$, increase in energy increases the potential. Thus the potential of the system increases with increase in plasma parameter $k$. The plots of $V_{eff}$ in Fig.(\ref{8}) displays both the co-rotating and counter rotating orbits. The co-rotating (prograde) orbits are characterised by $E>0$ and $L_{\phi}>0$ with respect to black hole spin $a>0$. The reverse happens in case of counter rotating (retrograde) orbits with $E>0$ and $L_{\phi}<0$ with respect to black hole spin $a>0$. We observe that the unstable photon orbit radius of counter rotating ($r_{p2}$) are greater than that of co-rotating orbits ($r_{p1}$) as can be seen from the maxima of the potential. This implies that the corotating orbits are near to the black hole than the counter rotating ones.

\section{Shadow radius $R_s$ and constraints from the M$87^*$ observational data}\label{observation}
In this section, we compute one of the major observables of black hole shadow, which is the black hole shadow radius ($R_s$). We do this by following the approach given in \cite{5b} which considers a reference circle to estimate the shadow radius. The definition of $R_s$ reads (in terms of the celestial coordinates)
\begin{eqnarray}
R_s = \frac{(\alpha_t -\alpha_r)^2+\beta_t^2}{2|\alpha_r-\alpha_t|}
\end{eqnarray}
where the silhouette of the shadow coincides with the reference circle at three different coordinates, the top coordinate ($\alpha_t,\beta_t$), the bottom coordinate ($\alpha_b,\beta_b$) and the right coordinate ($\alpha_r,0$). It is to be noted that the shadow radius $R_s$ is related to the angular diameter of the shadow as \cite{5c}
\begin{eqnarray}
\theta_d = 2 \frac{R_s}{d}
\end{eqnarray}
where $d$ represents the distance of $M87^*$ black hole from earth ($d=16.8$ Mpc). From the $EHT$ observations, the parameter $\theta_d$ is estimated to be $42\pm3$ $\mu a s$ or in radian $(0.20325\pm0.0146) \times 10^{-9}$ rad \cite{2}. We now try to constraint the PFDM parameter ($\frac{\chi}{M}$) and the plasma parameter ($k$) by confronting the theoretically estimated value of $\theta_d$ with the observational data. In the subsequent analysis, we set the charge parameter $\frac{Q}{M}=0$ for the sake of simplicity.\\
\noindent We first consider the Kerr limit ($\frac{\chi}{M} \rightarrow 0, k \rightarrow 0, \frac{Q}{M} \rightarrow 0$) of our black hole solution. We observe that for the variation of the spin parameter $\frac{a}{M} \in [0,1]$, the value of the angular diameter $\theta_d$ is obtained to be $\theta_d \in [0.1925\times10^{-9}, 0.1929\times10^{-9}]$ (in radian). This in turn means that all the values of the spin parameter $\frac{a}{M}$ produces angular diameter ($\theta_d$)  are allowed from the observational point of view.\\
Now incorporating the homogeneous plasma background ($h=0$) with $\frac{Q}{M}=0$, $\frac{\chi}{M}=0$, we try to constraint the value of the plasma parameter $k$ for various values of $\frac{a}{M}$. We obtain the possible range of $k$ from the observational constraint $\theta_d \in (0.20325\pm0.0146) \times 10^{-9}$ radian. The results are displayed in Table\ref{TabNew1}.\\
\begin{table}[h]
	\centering
	\begin{tabular}{|c|c|c|} 	 	
		\multicolumn{3}{c}{}\\
		\hline
		$\frac{a}{M}$ & $k_{lower}$ & $k_{upper}$ \\
		\hline
		0.1 & 0.0 & 0.976\\ 
		\hline
		0.2 & 0.0 & 0.909\\
		\hline
		0.3 & 0.0 & 0.805\\
		\hline
		0.4 & 0.0 & 0.677\\
		\hline
		0.5 & 0.0 & 0.536\\
		\hline
		0.6 & 0.0 & 0.396\\
		\hline
		0.7 & 0.0 & 0.266\\
		\hline
		0.8 & 0.0 & 0.154\\
		\hline
		0.9 & 0.0 & 0.067\\
		\hline
		1.0 & 0.0 & 0.020\\ 
		\hline
	\end{tabular}
	\caption{\footnotesize The results show the upper bound on the value of the plasma parameter $k$ ($0 \leq k \leq k_{upper}$) at fixed value of the spin parameter which results in the allowed value of $\theta_d$. (We set $\frac{\chi}{M} = 0, \frac{Q}{M} = 0$).}
	\label{TabNew1}
\end{table}
\noindent From Table\eqref{TabNew1} we note the range of allowed values of the plasma parameter $k$ at different values of the spin parameter $\frac{a}{M}$. It is to be noted that with increasing value of the spin parameter $\frac{a}{M}$, the allowed range for plasma parameter $k$ decreases. Further, we observe that when we consider the inhomogeneous plasma background ($h=1$), the resulting value of $\theta_d$ lies outside the estimated range of $\theta_d$. This is true for each and every value of the spin parameter $\frac{a}{M}$. Next we consider the PFDM black hole solution (with $\frac{Q}{M}\rightarrow 0$) in the homogeneous plasma background. The following Table represents our observation
\begin{table}[h]
	\centering
	\begin{tabular}{|c|c|c|} 	 	
		\multicolumn{3}{c}{}\\
		\hline
		$\frac{\chi}{M}$ & $k_{lower}$ & $k_{upper}$ \\
		\hline
		0.010 & 0.000 & 0.904\\ 
		\hline
		0.015 & 0.000 & 0.903\\
		\hline
		0.020 & 0.770 & 0.901\\
		\hline
		0.025 & 0.875 & 0.899\\
		\hline
	\end{tabular}
	\caption{\footnotesize The results show the lower and upper value of the plasma parameter $k$ at fixed value of the spin parameter, in presence of the PFDM parameter. (We set $ \frac{Q}{M} = 0$ and $\frac{a}{M}=0.2$).}
	\label{TabNew2}
\end{table} 

\noindent From Table\eqref{TabNew2}, we note that the allowed range of values for the PFDM parameter is $\frac{\chi}{M} \in [0, 0.025]$. We also observe the allowed range of values for $k$ corresponding to a fixed value of the PFDM parameter.

\noindent Next we consider the general case of plasma frequency $\omega_p (r, \theta)$. In this case, the plasma frequency is given by $\omega_p ^2 (r, \theta)  = \frac{f_r (r) + f_{\theta} (\theta)}{r^2 + a^2 \cos^2 \theta}$. We wish to constraint the $PFDM$ parameter ($\frac{\chi}{M}$) and the plasma parameter $\Big(\frac{\omega_c}{\omega_0}\Big)^2$ from the observed value of angular shadow ($\theta_d$). We consider the two cases studied in section \ref{I}. The first one is $f_r(r)=\omega_c ^2 \sqrt{r}$ and $f_{\theta} (\theta)=0$ and the other one $f_r(r)=0$ and $f_{\theta} (\theta)=\omega_c ^2 (1 + 2 \sin^2 \theta)$. The Tables below show the corresponding ranges of $\frac{\chi}{M}$ compatible with the ranges of $\Big(\frac{\omega_c}{\omega_0}\Big)^2$.
\begin{table}[H]
	\centering
	\begin{tabular}{|c|c|c|} 	 	
		\multicolumn{3}{c}{}\\
		\hline
		$\frac{\chi}{M}$ & $\Big(\frac{\omega_c}{\omega_0}\Big)_{lower} ^2$ & $\Big(\frac{\omega_c}{\omega_0}\Big)_{upper} ^2$ \\
		\hline
		0.001 & 0.0 & 0.489\\ 
		\hline
		0.002 & 0.0 & 0.387\\
		\hline
		0.003 & 0.0 & 0.293\\
		\hline
		0.004 & 0.0 & 0.205\\
		\hline
		0.005 & 0.0 & 0.121\\
		\hline
		0.006 & 0.0 & 0.039\\
		\hline
	\end{tabular}
	\caption{\footnotesize Table showing the accessible range of $\Big(\frac{\omega_c}{\omega_0}\Big) ^2$ with $f_r (r)=\omega_c ^2 \sqrt{r}$ and $f_{\theta} (\theta)=0$ for various values of $\frac{\chi}{M}$. The results show the upper bound on the value of the plasma parameter $\Big(\frac{\omega_c}{\omega_0}\Big) ^2$ at fixed value of the spin parameter ($\frac{a}{M}=0.5$) which results in the allowed value of $\theta_d$. (We set  $\frac{Q}{M} = 0$).}
	\label{TabNew3}
\end{table}

\begin{table}[h]
	\centering
	\begin{tabular}{|c|c|c|} 	 	
		\multicolumn{3}{c}{}\\
		\hline
		$\frac{\chi}{M}$ & $\Big(\frac{\omega_c}{\omega_0}\Big)_{lower} ^2$ & $\Big(\frac{\omega_c}{\omega_0}\Big)_{upper} ^2$ \\
		\hline
		0.001 & 0.0 & 0.282\\ 
		\hline
		0.002 & 0.0 & 0.223\\
		\hline
		0.003 & 0.0 & 0.168\\
		\hline
		0.004 & 0.0 & 0.117\\
		\hline
		0.005 & 0.0 & 0.069\\
		\hline
		0.006 & 0.0 & 0.022\\
		\hline
	\end{tabular}
	\caption{\footnotesize Table showing the accessible range $\Big(\frac{\omega_c}{\omega_0}\Big) ^2$ with $f_r (r)=0$ and $f_{\theta} (\theta)=\omega_c ^2 (1+ 2 \sin^2 \theta)$ for various values of $\frac{\chi}{M}$. The results show the upper bound on the value of the plasma parameter $\Big(\frac{\omega_c}{\omega_0}\Big) ^2$ at fixed value of the spin parameter ($\frac{a}{M}=0.5$) which results in the allowed value of $\theta_d$. (We set  $\frac{Q}{M} = 0$).}
	\label{TabNew4}
\end{table}

\noindent The above Tables \ref{TabNew3} and \ref{TabNew4} show that the allowed ranges of the $PFDM$ parameter $\frac{\chi}{M}$ is $\frac{\chi}{M} \in [0, 0.006]$ which implies the presence of a very small amount of dark matter in the vicinity of the black hole. Also the Tables entail that dark matter and plasma can coexist as depicted from the observational range of black hole shadow $\theta_d$. Besides, we observe that with the increase of $PFDM$ parameter $\frac{\chi}{M}$ from 0 to 0.006 the allowed range of plasma parameter $\Big(\frac{\omega_c}{\omega_0}\Big) ^2$ gradually decreases and vanishes from $\frac{\chi}{M}=0.007$.

\section{Summary and conclusion}\label{sec7}
After running through the analysis in details, we now summarise our results. We consider a charged rotating black hole surrounded by perfect fluid dark matter ($PFDM$). We also immerse the system in plasma and consider no interaction between plasma and dark matter. 

\noindent We observe some unique characteristics of the black hole spacetime due to presence of $PFDM$.
From the analysis of the function $\Delta(r)$, we find that the outer event horizon ($r_{h+}$) decreases with increase in
PFDM parameter ($\chi$). The decrement proceeds untill $\chi<\chi_c$. At this point and beyond, we observe a
reverse nature where $r_{h+}$ increases with further increase in $\chi$. The same behaviour gets reflected in the
observed black hole shadow. The reason for such a fascinating observation can be assigned to the contribution of PFDM to
the mass of the black hole system. Here, we have two masses, $M$ for the original black hole and $M_0$ for the black hole
corresponding to PFDM. Till $\chi<\chi_c$, the black hole mass is M and gets inhibited by $M_0$. But just
at the point when $\chi=\chi_c$, the total mass is given by $M_0$ beyond which system mass increases with increase in $\chi$. Since the mass has a major effect in determining the event horizon of the black hole and thereby the
shadow, hence we obtain such results.

\noindent Then we move on to analyse the motion of null particles around the black hole. These particles can be rotating along the direction of black hole spin (co-rotating) or opposite (counter rotating) to it. We observed that those co-rotating orbits lie close to the black hole whereas, the counter rotating ones remain away from the black hole. We find their dependence on plasma parameter. The effect of plasma is independent of the influence of dark matter. We find that in case of homogeneous plasma distribution \Big($n=\sqrt{1-k}$\Big), the radius of co-rotating orbits decrease whereas that for the counter rotating orbits increase with increase in plasma parameter $k$. Also, we find that with increase in plasma parameter ($k$) in case of inhomogeneous plasma distribution \Big($n=\sqrt{1-\frac{k}{r}}$\Big), the increase in $k$ results in decrease of photon radius for both co-rotating and counter rotating orbits.

\noindent Then we obtain the null geodesics responsible for the formation of the black hole shadow. With the help of the geodesic equation(s), we obtain the celestial coordinates ($\alpha, \beta$). These two coordinates give the black hole shadow radius ($R_s$) as $R_s ^2 = \alpha^2 + \beta^2$. The shadow gets formed in the celestial plane ($\alpha - \beta$ plane). We plot the shadow and study the dependence of the black hole shadow on the black hole parameters ($a$, $Q$, $\chi$, $k$). From the plots, we find that the shadow gets rotated and deformed with increase in black hole spin ($a$). The deformation of black hole shadow occurs due to the rotational drag of the unstable photons by the black hole.  We also observe that increment in charge ($Q$) reduces the radius ($R_s$) and thereby the size of the black hole shadow. The reason for this is quite obvious. The black hole shadow is the image of the outer event horizon of the black hole. The radius of outer event horizon is given as $r_{h+} = M + \sqrt{M^2 - Q^2}$ in absence of $\chi$ and $k$. With increase in $Q$, $r_{h+}$ decreases and thereby $R_s$ decreases. This ultimately reduces the size of the black hole shadow. We analyse this in case of inhomogeneous plasma distribution with \Big($n=\sqrt{1-k}$\Big).

\noindent Next we analyse the effect of the plasma medium on the black hole shadow. The shadow is formed by the light rays encircling the black hole in unstable photon orbits. If these light rays move through vacuum, they have no deviation. But if they move through plasma media, they get deviated due to variation in  frequency. The black hole shadow is formed in the celestial plane with coordinates $\alpha$ and $\beta$. We considered the general case where the refractive index ($n(r, \theta)$) and plasma frequency ($\omega_p (r, \theta)$) both depends on $r$ and $\theta$. We have considered a different value for the spin, $a=0.4$ in contrast to that considered in \cite{54} which is $a=0.999$ (extremal case). We found that with only radial variation 
$f_r (r) = \omega_c ^2 \sqrt{M^3 r}$ and $f_{\theta}(\theta)=0$, the shadow plots reduce in size with increase in plasma parameter $\omega_c$. Similar nature is observed in case of $\theta$ variation with  $f_r (r) = 0$ and $f_{\theta}(\theta)= \omega_c ^2 M^2 \Big(1 + 2 \sin^2 \theta\Big)$ with the shadow size decreasing with increase of plasma parameter $\omega_c$. 

\noindent Next we consider cases with refractive index of the form \Big($n=\sqrt{1-\frac{k}{r}}$\Big) (inhomogeneous) and \Big($n=\sqrt{1-k}$\Big) (homogeneous). Analysing the plots, we find that the extreme right of $\alpha$ axis corresponds to the radius of counter rotating orbits ($r_{p2}$) whereas that on the extreme left corresponds to the co-rotating orbits ($r_{p1}$). The variation of the radius of these orbits with plasma gets reflected in the black hole shadow. In particular, we observe that $r_{p1}$ decreases with increase in the plasma parameter for fixed value of the $PFDM$ parameter for both  inhomogeneous and homogeneous plasma. However, $r_{p2}$ increases with decreases in the plasma parameter ($k$) for a fixed $PFDM$ parameter ($\chi$) for inhomogeneous plasma, and increases for homogeneous plasma.

\noindent We have also analysed the effective potential ($V_{eff}$) and have found that it significantly depends on the plasma parameter ($k$). The maxima of the potential ($V_{eff}$) corresponds to the radius of the unstable photon orbits. These maxima's shift towards left for both homogeneous and inhomogeneous plasma distribution. Besides, we also find that the peak of the effective potential increases with increase in plasma parameter $k$. The reason for such an increment can be related to the fact that due to presence of plasma, the interaction energy of the total system increases and hence the potential ($V_{eff}$) of the system increases.\\
Finally, we compute the shadow radius $R_s$ and the angular shadow radius $\theta_d$. From our theoretical results, we constraint the plasma parameter $k$ as well as $\Big(\frac{\omega_c}{\omega_0}\Big) ^2$ with the $PFDM$ parameter $\frac{\chi}{M}$ by comparing the obtained values of $\theta_d$ with that observed from the $M87^*$ supermassive black hole data.

\section*{Acknowledgments}

A.D. would like to acknowledge the support of S.N. Bose National Centre for Basic Sciences for Senior Research Fellowship. A.S. acknowledges the financial support by Council of Scientific and Industrial Research (CSIR, Govt. of India). 
The authors would also like to thank the referees for useful comments.


\begin{thebibliography}{99}
\bibitem{2}   Event Horizon Telescope Collaboration, \emph{First M87 Event Horizon Telescope Results. I. The Shadow of
the Supermassive Black Hole}, \emph{The Astrophysical Journal Letters} \textbf{875} (2019) L1.
\bibitem{3} J. L. Synge, \emph{The escape of photons from gravitationally intense stars}, \emph{Mon. Not. R. Astron. Soc.} \textbf{131} (1966) 463-466.
\bibitem{4} J.P. Luminet, \emph{Image of a Spherical black hole with Thin Accretion disk}, \emph{Astron. Astrophys} \textbf{75} (1979) 228-235.
\bibitem{5} J.M. Bardeen, \emph{Timelike and null geodesics in the Kerr metric}, \emph{Black holes (Les astres occlus)} (1973) 215-239.
\bibitem{6} K. Hioki and K.I. Maeda, \emph{Measurement of the Kerr spin parameter by observation of a compact object’s shadow}, \emph{Phys. Rev. D} \textbf{80} (2009) 024042.
\bibitem{7} A de Vries, \emph{The apparent shape of a rotating charged black hole, closed photon orbits and the bifurcation set $A_4$}, \emph{Class. Quantum Grav.} \textbf{17} (2000) 123.
\bibitem{8} A. Grenzebach, V. Perlick and C. L\"{a}mmerzahl, \emph{Photon regions and shadows of accelerated black holes}, \emph{International Journal of Modern Physics D} \textbf{24} (2015) 9, 1542024.
\bibitem{9} A. \"{O}vg\"{u}n,  \.{I}. Sakallı and J. Saavedraa, \emph{Shadow cast and deflection angle of Kerr-Newman-Kasuya spacetime}, \emph{JCAP} \textbf{10} (2018) 041.
\bibitem{10} S. W. Wei, Y. C. Zou, Y. X. Liu, R. B. Mann, \emph{Curvature radius and Kerr black hole shadow}, \emph{JCAP} \textbf{08} (2019) 030.
\bibitem{11} A. Abdujabbarov, M. Amir, B. Ahmedov and S. G. Ghosh, \emph{Shadow of rotating regular black holes}, \emph{Phys. Rev. D} \textbf{93} (2016) 104004.
\bibitem{12} M. Wang, S. Chen, J. Jing, \emph{Shadow casted by a Konoplya-Zhidenko rotating non-Kerr black hole}, \emph{JCAP} \textbf{10} (2017) 051.
\bibitem{13} A. K. Mishra, S. Chakraborty, S. Sarkar, \emph{Understanding photon sphere and black hole shadow in dynamically evolving spacetimes}, \emph{Phys. Rev. D} \textbf{99} (2019) 104080.
\bibitem{14} R. Kumar, B. P. Singh, S. G. Ghosh, \emph{Shadow and deflection angle of rotating black hole in asymptotically safe gravity}, \emph{Annals Phys.} \textbf{120} (2020) 168252.
\bibitem{15} R. A. Hennigar, M. B. J. Poshteh, and R. B. Mann, \emph{Shadows, signals, and stability in Einsteinian cubic gravity}, \emph{Phys. Rev. D} \textbf{97} (2018) 064041.
\bibitem{16} H. Khodabakhshi, A. Giaimo, R. B. Mann, \emph{Einstein Quartic Gravity: Shadows, Signals, and Stability}, \emph{Phys. Rev. D} \textbf{102} (2020) 044038.
\bibitem{17} M. Amir, S. G. Ghosh, \emph{Shapes of rotating nonsingular black hole shadows}, \emph{Phys. Rev. D} \textbf{94} (2016) 024054.
\bibitem{18} R. Shaikh, \emph{Black hole shadow in a general rotating spacetime obtained through Newman-Janis algorithm}, \emph{Phys. Rev. D} \textbf{100} (2019) 024028.
\bibitem{19} E. Contreras, \'{A}. Rinc\'{o}n, G. Panotopoulos, P. Bargue$\tilde{n}$o, and B. Koch, \emph{Black hole shadow of a rotating scale-dependent black hole}, \emph{Phys. Rev. D} \textbf{101} (2020) 064053.
\bibitem{20} H. L\"{u} and H. D. Lyu, \emph{Schwarzschild black holes have the largest size}, \emph{Phys. Rev. D} \textbf{101} (2020) 044059.
\bibitem{21} X. H. Feng, H. L\"{u}, \emph{On the size of rotating black holes}, \emph{Eur. Phys. J. C} \textbf{80} (2020) 551.
\bibitem{22} R. Kumar, S. G. Ghosh, and A. Wang, \emph{Gravitational deflection of light and shadow cast by rotating Kalb-Ramond black holes}, \emph{Phys. Rev. D} \textbf{101} (2020) 104001.
\bibitem{23}  L. Ma,  H. L\"{u}, \emph{Bounds on photon spheres and shadows of charged black holes in Einstein-Gauss-Bonnet-Maxwell gravity}, \emph{Phys. Letters B} \textbf{807} (2020) 135535.
\bibitem{24} R. A. Konoplya and A. Zhidenko, \emph{Analytical representation for metrics of scalarized Einstein-Maxwell black holes and their shadows}, \emph{Phys. Rev. D} \textbf{100} (2019) 044015.
\bibitem{25} R. Kumar, S. G. Ghosh, and A. Wang, \emph{Shadow cast and deflection of light by charged rotating regular black holes}, \emph{Phys. Rev. D} \textbf{100} (2019) 124024.
\bibitem{a} S. Dastan, R. Saffari, S. Soroushfar, \emph{Shadow of a Kerr-Sen dilaton-axion Black Hole}, 	arXiv:1610.09477 [gr-qc].
\bibitem{b} A. Das, A. Saha, S. Gangopadhyay, \emph{Shadow of charged black holes in Gauss–Bonnet gravity}, \emph{Eur.Phys.J.C } \textbf{80} (2020) 3, 180.
\bibitem{c} A. Saha, S. M. Modumudi, S. Gangopadhyay, \emph{Shadow of a noncommutative geometry inspired Ayón Beato García black hole}, \emph{Gen.Rel.Grav.} \textbf{50} (2018) 8, 103.
\bibitem{d} L. Amarilla, E. F. Eiroa, \emph{Shadow of a rotating braneworld black hole}, \emph{Phys. Rev. D} \textbf{85} (2012) 064019.
\bibitem{e} L. Amarilla, E. F. Eiroa, \emph{Shadow of a Kaluza-Klein rotating dilaton black hole}, \emph{Phys. Rev. D} \textbf{87} (2013) 044057.
\bibitem{f} A. Grenzebach, V. Perlick, C. L\"{a}mmerzahl, \emph{Photon Regions and Shadows of Kerr-Newman-NUT Black Holes with a Cosmological Constant}, \emph{Phys. Rev. D} \textbf{89} (2014) 124004.
\bibitem{26} L. Amarilla, E. F. Eiroa, G. Giribet, \emph{Null geodesics and shadow of a rotating black hole in extended Chern-Simons modified gravity}, \emph{Phys. Rev. D} \textbf{81} (2010) 124045.
\bibitem{27} S. Dastan, R. Saffari, S. Soroushfar, \emph{Shadow of a Charged Rotating Black Hole in f(R) Gravity}, 	arXiv:1606.06994 [gr-qc].
\bibitem{28} R. Kumar, B. P. Singh, Md. S. Ali, S. G. Ghosh, \emph{Rotating black hole shadow in Rastall theory}, 	arXiv:1712.09793 [gr-qc].
\bibitem{29} T. Vetsov, G. Gyulchev, S. Yazadjiev, \emph{Shadows of Black Holes in Vector-Tensor Galileons Modified Gravity}, 	arXiv:1801.04592 [gr-qc].
\bibitem{30} H. M. Wang, Y. M. Xu, S. W. Wei, \emph{Shadows of Kerr-like black holes in a modified gravity theory}, \emph{JCAP} \textbf{03} (2019) 046.
\bibitem{31} T. Zhu, Q. Wu, M. Jamil, K. Jusufi, \emph{Shadows and deflection angle of charged and slowly rotating black holes in Einstein-Æther theory}, \emph{Phys. Rev. D} \textbf{100} (2019) 044055.
\bibitem{32} A. \"{O}vg\"{u}n, \.{I}. Sakallı, J. Saavedra and C. Leiva, \emph{Shadow cast of noncommutative black holes in Rastall gravity}, \emph{Modern Phys. Letters A} \textbf{35} (2020) 20, 2050163.

\bibitem{33} U. Papnoi, F. Atamurotov, S. G. Ghosh, B. Ahmedov, \emph{Shadow of five-dimensional rotating Myers-Perry black hole}, \emph{Phys. Rev. D} \textbf{90} (2014) 024073.
\bibitem{34} A. Abdujabbarov, F. Atamurotov, N. Dadhich, B. Ahmedov, Z. Stuchl\'{i}k \emph{Energetics and optical properties of 6-dimensional rotating black hole in pure Gauss-Bonnet gravity}, \emph{Eur. Phys. J. C } \textbf{75} (2015) 399.
\bibitem{35} B. P. Singh, S. G. Ghosh, \emph{Shadow of Schwarzschild–Tangherlini black holes}, \emph{Annals of Physics} \textbf{395} (2018) 127-137.

\bibitem{36} V. Perlick, O. Yu. Tsupko, G. S. Bisnovatyi-Kogan, \emph{Black hole shadow in an expanding universe with a cosmological constant}, \emph{Phys. Rev. D} \textbf{97} (2018) 104062.
\bibitem{37} P. C. Li, M. Guo, B. Chen, \emph{Shadow of a Spinning Black Hole in an Expanding Universe}, \emph{Phys. Rev. D} \textbf{101} (2020) 084041.

\bibitem{38} P. J. E. Peebles and B. Ratra, \emph{The cosmological constant and dark energy}, \emph{Rev. Mod. Phys} \textbf{75} (2003) 559.
\bibitem{39} K. Koyama, \emph{The cosmological constant and dark energy in braneworlds}, \emph{Gen Relativ Gravit} \textbf{40} (2007) 421-450.

\bibitem{40} J. Zhang, X. Zhang, H. Liu, \emph{Agegraphic dark energy as a quintessence}, \emph{Eur. Phys. J. C} \textbf{54} (2008) 303-309.
\bibitem{41} S. Tsujikawa, \emph{Quintessence: a review}, \emph{ Class. Quantum Grav.} \textbf{30} (2013) 214003.

\bibitem{42} V. C. Rubin, Jr. W. K. Ford, and N. Thonnard, \emph{Rotational properties of 21 SC galaxies with a large range of luminosities and radii, from NGC 4605 (R=4kpc) to UGC 2885 (R=122kpc)}, \emph{Astro. Phys. J.} \textbf{238} (1980) 471.
\bibitem{43} F. Zwicky, \emph{The redshift of extragalactic nebulae}, \emph{Helv. Phys.Acta.} \textbf{6} (1933) 110.

\bibitem{001} S. Tulin and H. B. Yu, \emph{Dark Matter Self-interactions and Small Scale Structure}, \emph{ Phys. Rep.} \textbf{730} (2018) 1.
\bibitem{44} V. V. Kiselev, \emph{Quintessence and black holes}, \emph{Class. Quantum Grav.} \textbf{20} (2003) 6, 1187.
\bibitem{45} V. V. Kiselev, \emph{Quintessential solution of dark matter rotation curves and its simulation by extra dimensions}, arXiv: 0303031 [gr-qc].

\bibitem{46} M. H. Li, K. C. Yang, \emph{Galactic dark matter in the phantom field}, \emph{Class. Quantum Grav.} \textbf{86} (2012) 123015.
\bibitem{47} Z. Xu, X. Hou, J. Wang, \emph{Kerr-anti-de Sitter/de Sitter black hole in perfect fluid dark matter background}, \emph{Class. Quantum Grav.} \textbf{35} (2018) 115003.
\bibitem{48} Z. Xu, X. Hou, J. Wang, Y. Liao, \emph{Perfect Fluid Dark Matter Influence on Thermodynamics and
Phase Transition for a Reissner-Nordstrom-Anti-de Sitter Black Hole}, \emph{Adv. in High Energy Phys.} (2019) 2434390.
\bibitem{49} S. Haroon, M. Jamil, K. Jusufi, K. Lin and R. B. Mann, \emph{Shadow and deflection angle of rotating black holes in perfect fluid
dark matter with a cosmological constant}, \emph{Phys. Rev. D} \textbf{99} (2019) 044015.
\bibitem{50} X. Hou, Z. Xu, J. Wang, \emph{Rotating black hole shadow in perfect fluid dark
matter}, \emph{JCAP} \textbf{12} (2018) 040.
\bibitem{1}  A. Das, A. Saha and S. Gangopadhyay, \emph{Investigation of the circular geodesics in a rotating charged black hole in presence of perfect fluid dark matter}, \emph{Class. Quantum Grav.} \textbf{38} (2021) 6, 065015.
\bibitem{x}  Y. Heyderzade and F. Darabi, \emph{Black hole solutions surrounded by perfect fluid in Rastall theory}, \emph{Phys. Letts. B} \textbf{771} (2017) 365-373.
\bibitem{68} S. Nampalliwar, S. Kumar, K. Jusufi, Q. Wu, M. Jamil and P. Salucci, \emph{Modeling the Sgr A$^{*}$ Black Hole Immersed in a Dark Matter Spike}, \emph{The Astrophysical Journal} \textbf{916} (2021), 116.
\bibitem{1aa} V. Perlick and O. Yu. Tsupko, \emph{Calculating black hole shadows: review of analytical studies}, \emph{Phys. Reps.} \textbf{947} (2022) 1-39.
\bibitem{67} First M87 Event Horizon Telescope Results. VIII. Magnetic Field Structure near The Event Horizon, \emph{The Astrophysical Journal Letters} \textbf{910} (2021) L13.
\bibitem{51} V. Perlick, O. Yu. Tsupko, and G. S. Bisnovatyi-Kogan, \emph{Influence of a plasma on the shadow of a spherically symmetric black hole}, \emph{Phys. Rev. D} \textbf{92} (2015) 104031.
\bibitem{52} F. Atamurotov, B. Ahmedov, and A. Abdujabbarov, \emph{Optical properties of black holes in the presence of a plasma: The shadow}, \emph{Phys. Rev. D} \textbf{92} (2015) 084005.
\bibitem{53} M. Sharif, S. Iftikhar , \emph{Shadow of a charged rotating non-commutative black hole}, \emph{Eur. Phys. J. C} \textbf{76} (2016) 630.
\bibitem{54} V. Perlick, O. Yu. Tsupko, \emph{Light propagation in a plasma on Kerr spacetime: Separation of the Hamilton-Jacobi equation and calculation of the shadow}, \emph{Phys. Rev. D} \textbf{95} (2017) 104003.
\bibitem{55} G. S. Bisnovatyi-Kogan, O. Yu. Tsupko, \emph{Gravitational Lensing in Presence of Plasma: Strong Lens Systems, Black Hole Lensing and Shadow}, \emph{Universe} \textbf{3} (2017) 3, 57.
\bibitem{56} A. Abdujabbarov, B. Toshmatov, Z. Stuchl\'{i}k and B. Ahmedov, \emph{Shadow of the rotating black hole with quintessential energy in the presence of plasma}, \emph{IJMPD} \textbf{26} (2017) 6, 1750051.
\bibitem{57} Y. Huang, Y. P. Dong and D. J. Liu, \emph{Revisiting the shadow of a black hole in the presence of a plasma}, \emph{IJMPD} \textbf{27} (2018) 12, 1850114.
\bibitem{58} B. Ahmedov, B. Turimov, Z. Stuchl\'{i}k and A. Tursunov, \emph{Optical properties of magnetized black hole in plasma}, \emph{IJMPD: Conference Series} \textbf{49} (2019) 1960018.
\bibitem{59} G. Z. Babar, A. Z. Babar, F. Atamurotov, \emph{Optical properties of Kerr–Newman spacetime in the presence of plasma}, \emph{Eur. Phys. J. C} \textbf{80} (2020) 761.
\bibitem{60} A. Chowdhuri, A. Bhattacharyya, \emph{Shadow analysis for rotating black holes in the presence of plasma for an expanding universe},  \emph{Phys. Rev. D} \textbf{104} (2021) 064039.
\bibitem{61} C. Q. Liu, C. K. Ding, J. L. Jing, \emph{Effects of Homogeneous Plasma on Strong Gravitational Lensing of Kerr Black Holes}, \emph{Chin. Phys. Lett.} \textbf{34} (2017) 9, 090401.
\bibitem{1b} H. X. Zhang, Y. Chen, P. Z. He, Q. Q. Fan, and J. B. Deng, \emph{Bardeen black hole surrounded by perfect fluid dark matter}, \emph{Chin. Phys. C} \textbf{45} (2021) 5, 055103. 

\bibitem{62} R. A. Breuer, J. Ehlers, \emph{ Propagation of electromagnetic waves through magnetized plasmas in arbitrary gravitational fields}, \emph{Astron. Astrophys.} \textbf{96} (1981) 1-2, 293-295.
\bibitem{63} V. Perlick, \emph{Ray Optics, Fermat’s Principle, and Applications to General Relativity}, (Springer, Berlin, 2000).
\bibitem{64} J. L. Synge, \emph{Relativity. The General Theory}, (North-Holland, Amsterdam, 1960).
\bibitem{65} A. Rogers, \emph{Frequency-dependent effects of gravitational lensing within plasma}, \emph{Mon. Not. Royal Astron. Soc.} \textbf{451} (2015) 1, 17-25.
\bibitem{66} S. Chandrasekhar, \emph{The Mathematical Theory of Black Holes}, Oxford University Press, Oxford (1998).
\bibitem{5a} R. Shaikh, K. Pal, K. Pal and T. Sarkar, \emph{Constraining alternatives to the Kerr black hole}, \emph{Mon. Not. Royal Astron. Soc.} \textbf{506} (2021), 1229–1236.
\bibitem{5b} K. Hioki and K. Maeda, \emph{Measurement of the Kerr spin parameter by observation of a compact objects shadow},  \emph{Phys. Rev. D} \textbf{80} (2009) 024042.
\bibitem{5c} R. Kumar and S. G. Ghosh, \emph{Rotating black holes in 4D Einstein-Gauss-Bonnet
gravity and its shadow},  \emph{JCAP} \textbf{07} (2020) 053.
\end{thebibliography}
\end{document}